\documentclass[11pt, 5p, nopreprintline]{elsarticle}

\journal{TBA}

\biboptions{numbers,sort&compress}

\usepackage{libertine}
\usepackage{libertinust1math}
\usepackage{amsmath}                 
\usepackage{bbold}                   
\usepackage{graphicx}                
\usepackage{eurosym}                 
\usepackage{mathtools}               
\usepackage{url}                     
\usepackage{booktabs}                
\usepackage{epstopdf}                
\usepackage{xfrac}                   
\usepackage{tabularx}                
\usepackage{bm}                      
\usepackage{subfig}
\usepackage{subcaption}              
\usepackage{longtable}               
\usepackage{multirow}                
\usepackage{threeparttable}          
\usepackage{pdflscape}               
\usepackage[svgnames]{xcolor}    	 
\usepackage[export]{adjustbox}       
\usepackage[version=4]{mhchem}       
\usepackage{xurl} 				     
\usepackage[colorlinks]{hyperref}    
\usepackage[parfill]{parskip}        
\usepackage[nameinlink,sort&compress,capitalise,noabbrev]{cleveref} 
\usepackage[leftcaption,raggedright]{sidecap} 
\usepackage[prependcaption,textsize=footnotesize]{todonotes} 
\usepackage{siunitx}                 
\usepackage{csquotes} 				 
\usepackage[acronym, automake, nonumberlist]{glossaries-extra}
\usepackage{float}
\usepackage{lipsum}
\usepackage{mdframed}
\usepackage[resetlabels,labeled]{multibib} 
\newcites{S}{Supplementary References}
\bibliographystyleS{elsarticle-num}

\DeclareSIUnit\year{a}
\DeclareSIUnit{\tco}{t_{\ce{CO2}}}
\DeclareSIUnit{\sieuro}{\mbox{\euro}}

\graphicspath{ 						
	{imgs/},
	}

\newdefinition{rmk}{Remark}

\makeglossaries

\newacronym{iea}{IEA}{International Energy Agency}

\newacronym{lcoe}{LCOE}{Levelized Cost of Electricity}

\newacronym{egs}{EGS}{Enhanced Geothermal Systems}
\newacronym{orc}{ORC}{Organic Rankine Cycle}
\newacronym{dac}{DAC}{Direct Air Capture}
\newacronym{ocgt}{OCGT}{Open-Cycle Gas Turbine}
\newacronym{ccs}{CCS}{Carbon Capture and Storage}
\newacronym{capex}{CAPEX}{Capital Expenditure}

\begin{document}

\begin{frontmatter}

	\title{Market Integration Pathways for Enhanced Geothermal Systems in Europe}

	\author[uoe_address]{Lukas Franken} \ead{lukas.franken@ed.ac.uk}
    \author[tub_address]{Elisabeth Zeyen}
    \author[gla_address,ramboll_address]{Orestis Angelidis}
    \author[tub_address]{Tom Brown}
    \author[uoe_address]{Daniel Friedrich}

	\address[uoe_address]{School of Engineering, Institute for Energy Systems, University of Edinburgh, United Kingdom}
    \address[tub_address]{Department of Digital Transformation in Energy Systems, TU Berlin, Germany}
    \address[gla_address]{James Watt School of Engineering, University of Glasgow, United Kingdom}
    \address[ramboll_address]{Ramboll, 240 Blackfriars Rd, London, SE1 8NW, United Kingdom}

	\begin{abstract}
		Enhanced Geothermal Systems (EGS) can provide constant, reliable electricity and heat with minimal emissions, but high drilling costs and uncertain cost reductions leave their future unclear.  
We explore scenarios for the future adoption of \gls{egs} in a carbon-neutral,
multi-sector European energy system.  
We find that in a net-zero system, heat (co-)generating \gls{egs} at current cost can support 20–30$\,$GW\textsubscript{th} of capacity in Europe, primarily driven by district heating demands.
When drilling costs decrease by approximately 60\%, \gls{egs} becomes competitive in electricity markets, expanding its market opportunity by one order of magnitude. 
However, the spatially dispersed rollout of district heating contrasts with the confined overlap of high geological potential and low potential for other renewables, which conditions the competitiveness of electricity-generating \gls{egs}.
This results in a challenge where the majority of \gls{egs} market potential depends on pan-European technology learning for cost reductions, emphasising coordination is crucial in stakeholders' efforts to reduce \gls{egs} cost.

	\end{abstract}



\end{frontmatter}


\section*{Introduction}
\label{sec:intro}
Geothermal energy has the potential to serve as a firm and renewable energy source by harnessing the abundant heat from the subsurface \cite{iea2023outlook}. 
However, there is high uncertainty about future drilling costs, especially for \gls{egs}, resulting in varying projections of the geothermal contribution (1–4\%) to the future energy mix \cite{aghahosseini2020hot, doe2019geo, dalla2020scenarios}.

One of the primary differences between these estimates is the uncertain role of \gls{egs}.
In \gls{egs}, \enquote{Enhanced} refers to the synthetic increase in deep, hot rock's permeability to facilitate the heating of a working fluid for heat and/or electricity generation \cite{lu2018global}.
Unlike traditional geothermal systems, \gls{egs} is not restricted to regions with exceptional geological suitability but has broader applicability.
However, high drilling costs currently make \gls{egs} economically uncompetitive, despite its significant untapped potential \cite{aghahosseini2020hot}.
Should drilling costs decrease, \gls{egs} could play a crucial role in future energy supply.

Existing \gls{egs} projects often improve their economic viability by supplying heat in addition to electricity \cite{lu2018global}.
In these cases, thermal energy can be distributed to end users with high efficiencies 
compared to the much lower efficiencies achievable in electricity generation (approximately 10–20\%, depending on rock formation temperature) \cite{eyerer2020combined, tester2006future}.
This approach has enabled the development of over ten smaller \gls{egs} projects in Europe, producing between 1MW$_{\text{th}}$ and 11MW$_{\text{th}}$, all of which generate or co-generate heat \cite{jain2015maximum, genter2010contribution, gerard2006deep}.
Heat generation is particularly beneficial as it improves \gls{egs}' economic prospects \cite{tester2006future}, contributes low-carbon energy in the heating sector \cite{bertani2016geothermal}, better accommodates suboptimal geothermal fluid temperatures \cite{dipippo2012geothermal}, and is often supported by additional policy incentives \cite{eu2023geothermal}.

Existing modelling efforts evaluate the potential of \gls{egs} either by using geophysical data or modelling its market competitiveness in the whole energy system.
The estimates of \gls{egs} potential and levelized cost of electricity (\gls{lcoe}) vary significantly between studies.
A recent study estimates the global \gls{egs} potential to be around 1.5TW$_\text{el}$ at or below an \gls{lcoe} of approximately 150\euro/MWh$_\text{el}$.
Current technological capabilities allow for the extraction of 227GW$_\text{el}$ of this potential in Europe \cite{aghahosseini2020hot}.
In contrast, an earlier study estimated that the opportunity for \gls{egs} in Europe by 2050 could reach 522GW$_\text{el}$ at an \gls{lcoe} below 100\euro/MWh$_\text{el}$ \cite{limberger2014assessing}.
Longa et al. \cite{dalla2020scenarios} inserted this potential into an integrated assessment model and showed a cost-optimal \gls{egs} capacity of around 25GW$_\text{el}$ when competing with other technologies.
The exploration of \gls{egs} potential in an investment and dispatch power model for the United States West Coast highlights the importance of modelling \gls{egs} features that are only apparent at high temporal resolution, such as flexible operation \cite{ricks2024role}.
More recent research suggests that the economic prospects of \gls{egs} could be further enhanced by extracting lithium alongside with heat and/or power generation \cite{weinand2023low}.

Despite these efforts, existing literature fails to address some of the most crucial questions faced by industry stakeholders and policymakers looking to invest in or promote \gls{egs} deployment.

First, previous studies estimate the volume of \gls{egs} potential in Europe but do not provide an account of how that pathway could materialise.  
Without an assessment of the spatial distribution of market opportunities across varying levels of technological progress, stakeholders are left without the insight needed to support that development effectively.  
This shortcoming of past work arises from the computational burden associated with multi-year simulation horizons, which are commonly employed in integrated assessment models, such as those used, for instance, in Longa et al. \cite{dalla2020scenarios}.  
As a result, the intricate interaction between the spatial distribution of energy demand, renewable generation opportunities, and energy transport—constrained by the transmission grid—is overlooked.  
Thus, while an estimate of overall \gls{egs} potential has been established, the findings of \cite{dalla2020scenarios} need to be expanded to provide a detailed projection of how a European energy future with \gls{egs} might manifest.

\begin{figure}[t]
    \centering
    \includegraphics[width=0.95\columnwidth]{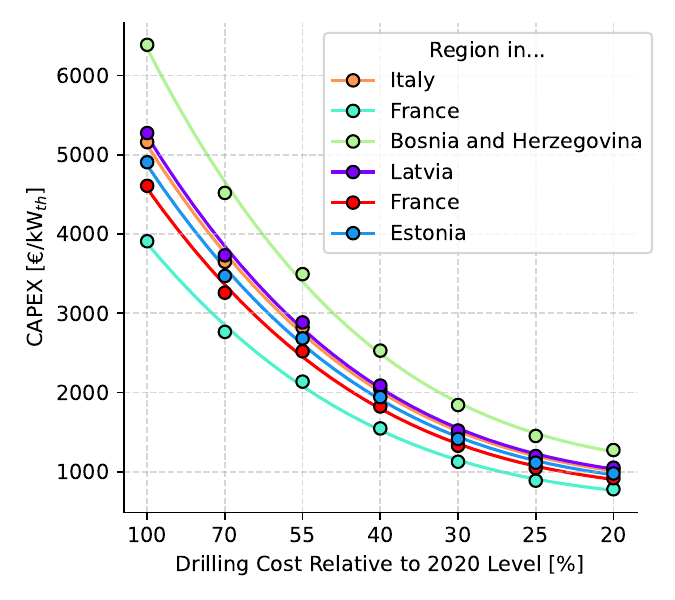}
    \caption{Scenarios of potential drilling cost reductions evaluated in the study, shown here for six out of 72 regions. All costs are assumed to decrease, however preserve the relative magnitude between regions, reflecting geological conditions. The scenarios cover the span of the cost reductions evaluated in \cite{aghahosseini2020hot}.}
    \label{fig:schematic_cost_reduction}
\end{figure}

Second, it also remains unclear which level of cost reduction would permit \gls{egs} to enter the European energy market.
The binary framing of cost development - either optimistic or conservative - applied in \cite{dalla2020scenarios} and \cite{ricks2024role} does not allow for this critical aspect of the analysis.
This is regrettable, since having a detailed account of market opportunity associated with different levels of \gls{egs} costs, both in terms of volume and spatial distribution, would indicate the ultimate value of potential technical advances.
In turn, missing this analysis leaves policymakers and investors uncertain about the likelihood that investing in \gls{egs} will pay off.

\begin{figure}[t]
    \centering
    \includegraphics[width=1\columnwidth]{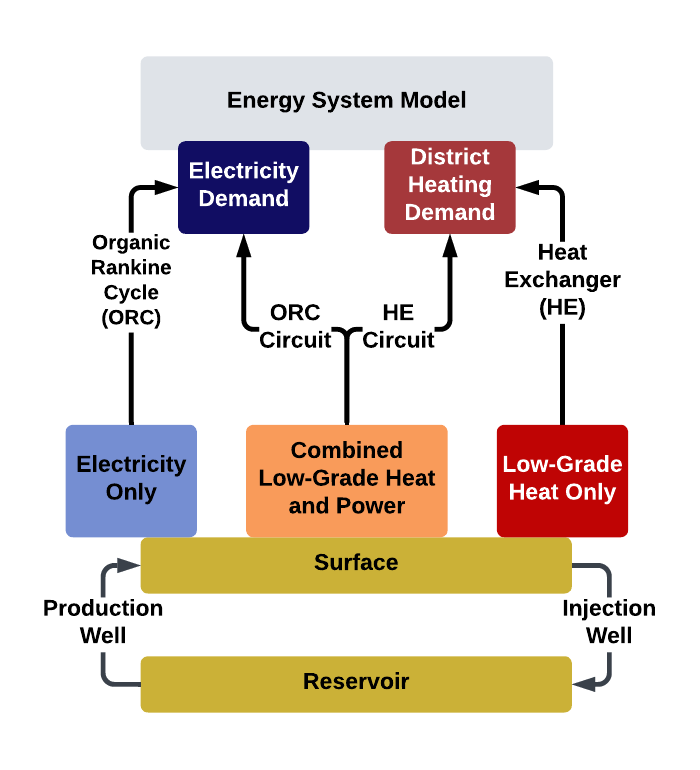}
    \caption{Flowchart of a simplified representation of the \gls{egs} model. Note that all model runs include \textit{one} of the three modes where \gls{egs} generates only electricity, combined low-grade heat and power or low-grade heat only. While different in the technologies they represent on a surface level, the subsurface modelling remains the same between them.}
    \label{fig:egs_flowchart}
\end{figure}

Here, we address these gaps by analysing \gls{egs}' future market opportunities in Europe with a high spatiotemporal resolution.
We incorporate \gls{egs} at varying levels of cost reduction in a sector-coupled, carbon-neutral European energy system model (Figure \ref{fig:schematic_cost_reduction}).
The analysis relies on the open-source model PyPSA-Eur (see Methods Subsection \ref{subsec:energy_system_model}), modified to include \gls{egs} in three distinct modes of generation where it generates: \textit{(i)} electricity only, \textit{(ii)} combined low-grade heat and power, or \textit{(iii)} low-grade heat only (Figure \ref{fig:egs_flowchart}, see Methods Subsection \ref{subsec:modelling_egs}).
In the latter two cases, the model is enabled to supply low-grade heat to district- and low-temperature industrial heat demands.
To capture the spatial distribution of \gls{egs}, we assume drilling costs based on existing estimates of geological suitability \cite{aghahosseini2020hot}.

This allows us to make several novel contributions:  \\
\textit{(1)} For each level of drilling cost reduction, we characterise the \gls{egs} market opportunity in terms of spatial distribution and the sectors supplied.  
In this analysis, we identify two phases of rollout: the first phase, where heat (co)generation is crucial for \gls{egs} to be competitive,  
and the second phase, where a tipping point in drilling costs is reached that enables \gls{egs} to compete directly with wind and solar, thereby unlocking a much larger market.  
\textit{(2)} We estimate the necessary learning rate to unlock larger markets.  
These rates would enable \gls{egs} rollout to become largely self-sustaining, with each step of cost reduction driving the subsequent one.  
\textit{(3)} We explore the cost-optimal spatial distribution of \gls{egs} deployment, and identify barriers and opportunities that can accelerate or hinder its rollout.  
These findings provide valuable context for both short-term \gls{egs} development and long-term investment and subsidy strategies that aim to support \gls{egs} deployment.

\label{sec:results}
%

\begin{figure}[t]
    \centering
    \includegraphics[width=0.95\columnwidth]{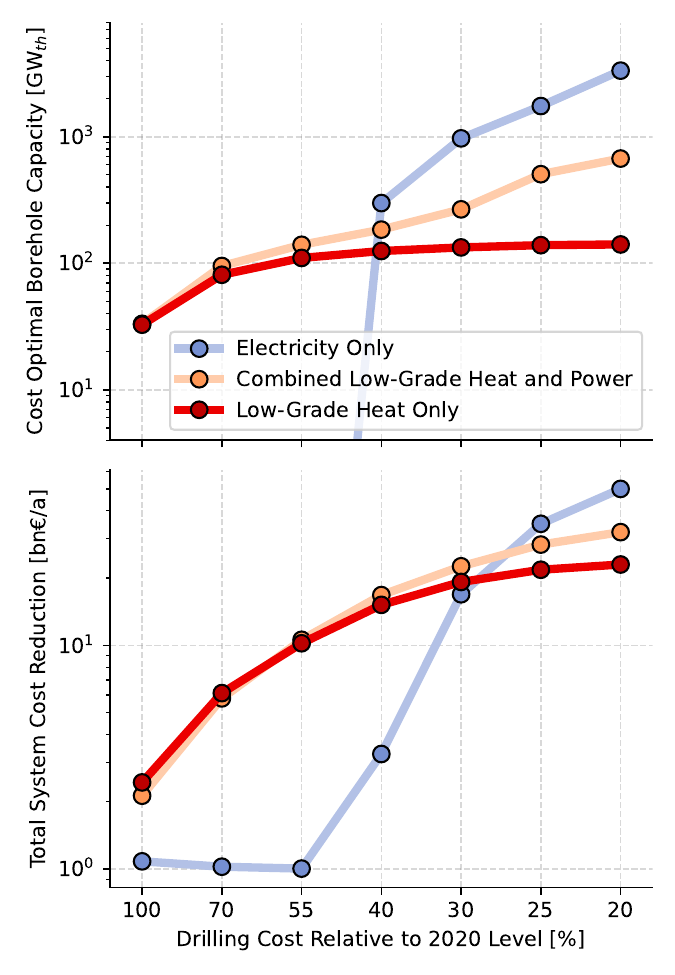}
    \caption{Demand curves for borehole thermal capacity of Enhanced Geothermal Systems in Europe and resulting total system cost reduction. The \textit{y}-axis in the upper plot refers to heat capacity.}
    \label{fig:main_plot}
\end{figure}

\begin{figure*}[t]
    \centering
    \includegraphics[width=\textwidth]{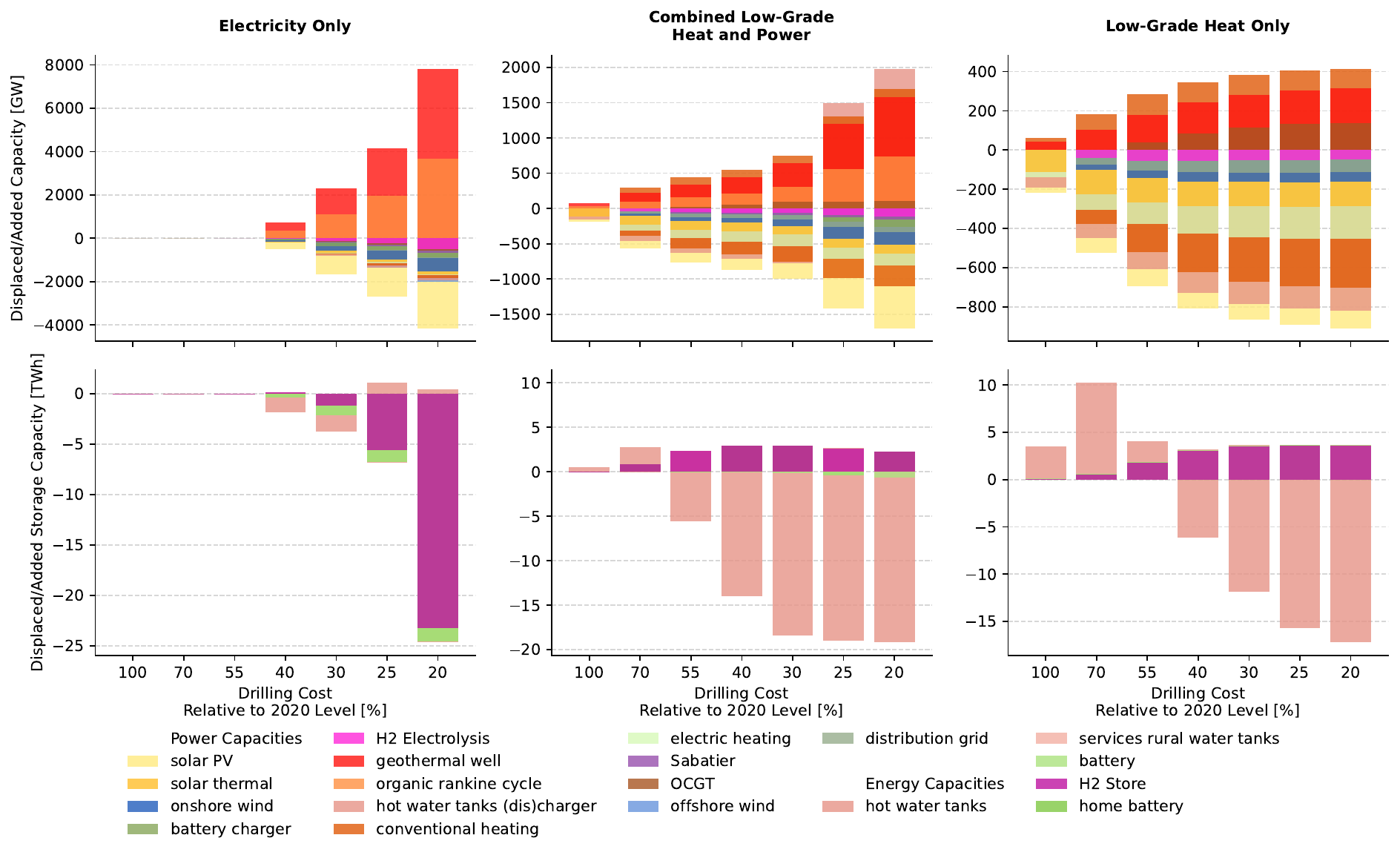}
    \caption{Changes of installed capacities for different \gls{egs} cost reduction levels. The \textit{top} row refers to power capacities, the \textit{bottom} row to energy storage capacities. Positive values are added, negative ones are subtracted relative to a counterfactual where \gls{egs} is not available.}
    \label{fig:displacement}
\end{figure*}

\begin{figure*}[t]
    \centering
    \includegraphics[width=\textwidth]{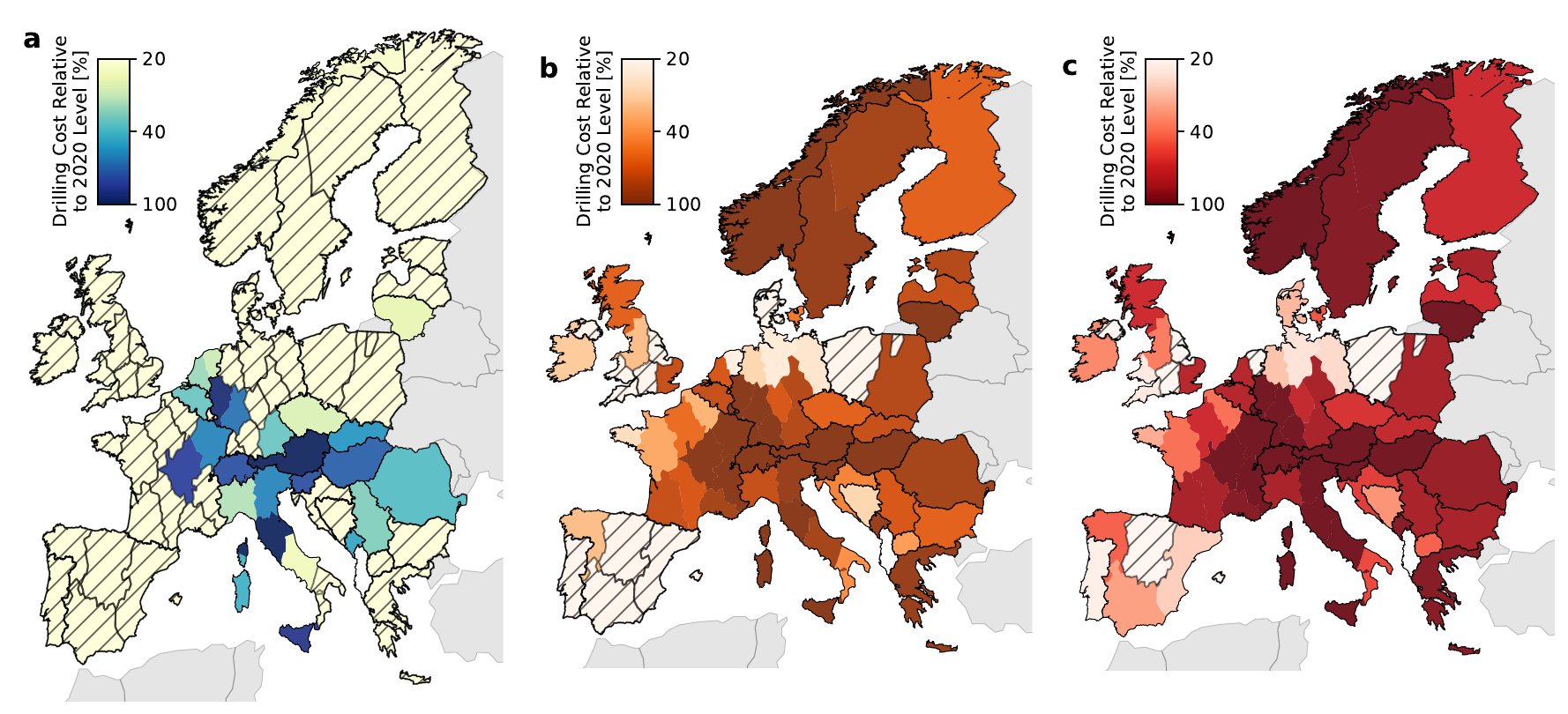}
    \caption{Cost reduction relative to today's estimated cost required for \gls{egs} to contribute 10\% of the respective demand per region. The maps show \gls{egs} generating \textbf{a} electricity, \textbf{b} combined low-grade heat and power and \textbf{c} low-grade heat only.}
    \label{fig:map_year}
\end{figure*}

\subsection*{\textbf{Heat Generating EGS as Market Entry}}

The cost competitiveness of \gls{egs} depends strongly on which energy service is provided (heat, electricity, or both) as well as the anticipated reduction in drilling cost (Figure \ref{fig:main_plot}).
Heat-generating \gls{egs}, at current costs, is already cost-competitive in a carbon-neutral energy system, leading to a reduction in overall system costs by approximately 0.5\% (2.5bn\euro/a).  
Co-generation of power and heat offers greater savings, with total system cost reductions ranging from 0.5\% to 3.6\% when compared to heat generation alone, which achieves savings of 0.5\% to 2.6\%. 
For \gls{egs} to become competitive as a source of electricity generation alone, drilling costs would need to decrease to 40\% of their current levels.
This scenario yields the largest total system cost reduction, with costs decreasing by about 50 billion\euro/a (5.74\%) when \gls{egs} costs are reduced to 20\% of their present levels.
However, the cost reduction per installed \gls{egs} unit is lowest for electricity generation, indicating that much of the installed electricity-generating \gls{egs} capacity is only marginally more cost-effective than alternative supply options.
In terms of borehole heat capacity, electricity generation peaks at approximately 4,000$\,$\\GW$\text{th}$, whereas co-generation peaks at 700$\,$GW$\text{th}$, and heat generation reaches a peak of 150$\,$GW$_\text{th}$.

As drilling costs reduce, the market for \gls{egs} grows.  
By connecting each level of cost reduction with a market opportunity, we can estimate the technology learning rate needed to reach the subsequent next level of cost reduction given the available market size (see Methods Subsection \ref{subsec:tech_learning}).  
We find that the learning rate to be around 20\% (Appendix Figure \ref{fig:learning_rate}).  
For low cost reductions, where learning opportunity is limited due to smaller heating demand and inaccessible electricity markets, a slightly higher learning rate of approximately 25\% is needed.  
As soon as electricity generation becomes the main application for \gls{egs}, the learning rate stabilises at around 20\%.

As \gls{egs} capacity gradually increases with cost reductions, it displaces growing volumes of alternative generation capacities, primarily wind, solar, and fossil-fuel heat generators (Figure \ref{fig:displacement}).  
In district heating, \gls{egs} at low to medium cost reduction changes the composition of supply drastically, replacing most solar thermal, gas, and biomass heating (Figures \ref{fig:percentage-displacement-chp}, \ref{fig:percentage-displacement-dh}).  
The freed-up biomass is reallocated to heating in industry, and the emission budget is used for OCGT electricity generation instead.  
For electricity-generating \gls{egs}, displacement is more gradual, with 5–11\% of solar and wind generation displaced at medium cost reductions and 16–40\% displaced at extreme cost reductions (Figure \ref{fig:percentage-displacement-elec}).  
For power capacities, the displacement varies in volume, but its proportionality remains largely consistent as \gls{egs} costs decrease. 
In terms of energy capacity, however, displacement depends on the level of cost reduction;  
for higher-cost \gls{egs}, storage is added to accommodate baseload generation, while for lower-cost \gls{egs}, it is removed in favour of operating \gls{egs} capacity at lower capacity factors.

For electricity generation at high cost reductions, the model uses the dispatchable nature of \gls{egs} by removing around 22$\,$TWh hydrogen storage capacity (equals 40\% of hydrogen storage capacity in \gls{egs}-free model), otherwise needed for balancing.
For heat (co-)gen-\\erating \gls{egs} at low cost reduction, hot water storages are initially added to the model to provide flexibility for the more base-load operation of \gls{egs}.
As cost reductions progress, these hot water storages are subsequently withdrawn along with solar thermal, leading to a decrease in installed thermal storage capacity of 20$\,$TWh (63\%) for \gls{egs} generating combined low-grade heat and power and 16$\,$TWh (58\%) for only low-grade heat.
Reductions in hot water tank energy capacity, alongside increases in (dis-)\\charging capacity, indicate a shift in the role of water tanks from providing short-term to offering more long-term flexibility.

Interestingly, a similar but opposite dynamic is observed for hydrogen storage in heat (co-)generating applications, where the model consistently installs additional H$_2$ storage capacity along with \gls{egs}.
Its increase appears to be favoured over the hydrogen supply chain from variable renewables with electrolysis in the \gls{egs}-free model.
The introduction of \gls{egs} also leads to substantial removal of fossil-fuel heating.
This reduction of emissions within the heating sector allows the model to favour more fossil fuel generation in the power sector, building up to nearly 140$\,$GW$_\text{th}$ (approximately 50\%) of additional \gls{ocgt} capacity, depending on the \gls{egs} cost reduction factor.

\begin{figure*}[t]
    \centering
    \includegraphics[width=1\textwidth]{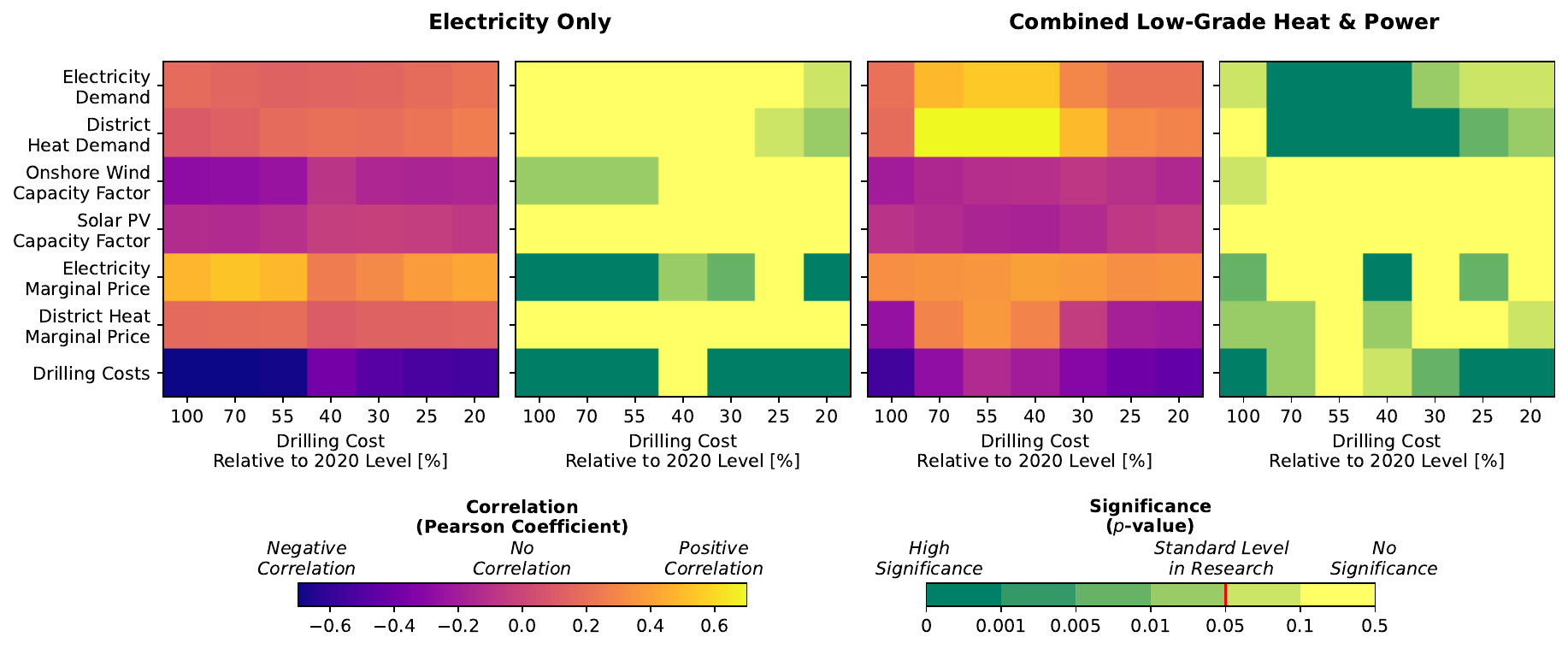}
    \caption{Correlation between cost-optimal borehole capacity and various features for electricity-generating EGS and combined low-grade heat and power EGS.
    Each column refers to one model run and evaluates both the correlations and the statistical significance of those correlations.
    For the rows representing electricity and district heat marginal price, the values refer to marginal prices in a system without EGS.
    The respective figure for heat-generating EGS is provided in the Figure \ref{fig:correlation_dh}.}
    \label{fig:correlation}
\end{figure*}

\subsection*{\textbf{Spatial Aspects of EGS Adoption}} \label{subsec:spatial}

The spatial distribution of \gls{egs} adoption results from the interplay of geological suitability, electricity prices in the incumbent system, and the distribution of energy demands.

This interplay turns out to vary between electricity and heat (co-)generating \gls{egs}.
For the former, both geological suitability and low renewable capacity factors are crucial for \gls{egs} to be competitive.
These conditions are met only in a small subset of regions, primarily around Central Europe, the Benelux countries, and the Northern Balkans.
In contrast, heat (co-)generating \gls{egs} is viable across a broader range of regions, even for low cost reductions. 
These regions include Central, Eastern, and Southeastern Europe, as well as Scandinavia.
This distribution is mainly driven by the magnitude of district heating demands and only to a lesser extent by geological suitability.

Echoing the findings of the previous section, electricity-generating \gls{egs} requires significant cost reductions to be viable (Figure \ref{fig:map_year}). 
However, the spatial concentration of electricity-generating \gls{egs} in a small subset of Central European regions is surprising, given the much larger cost-optimal \gls{egs} borehole capacities observed in the first part of the results. 
Even at an extremely low cost of 20\% relative to today's levels, most regions do not see substantial capacity expansion. 
Looking at the correlation between region's properties and their \gls{egs} uptake, we find that geological suitability is the main driver behind the concentrated adoption of electricity-producing \gls{egs} (Figure \ref{fig:correlation}). 
This is indicated by strong negative correlations between -0.6 and -0.3. 
However, a secondary predictor is the marginal price of electricity, with a correlation of around 0.4. 
High marginal prices are partially caused by low renewable capacity factors, a relationship also reflected in the minor negative correlations between \gls{egs} adoption and onshore wind capacity factors (around -0.2).

\gls{egs} with heat generation is installed in more regions, even at only minor cost reductions of less than 50\% of today's level. 
These regions include Central Europe, mainly around the Alps, parts of the Balkans and the Baltics, as well as Norway and Sweden. 
Once costs of around 40\% of today's levels are undercut, most regions adopt heat-using \gls{egs}; only Iberia, England, Northern Germany, and Western Poland remain mostly unfavourable to \gls{egs} even under high cost reductions (as indicated by hashed regions in Figure \ref{fig:map_year}). 
Here, the statistical assessment shows that the distribution of demands plays a significantly larger role in driving \gls{egs} market opportunity, with Pearson correlations between 0.4 and 0.6 at high significance (Figure \ref{fig:correlation}). 
Surprisingly, the correlation with solar capacity factors is negligible, despite solar thermal being one of the main technologies displaced by heat (co-)generating \gls{egs}. 
Likely, the influence of demand dominates this interaction, overshadowing the more subtle relationship with solar resources. 
For the spatial distribution of the predictors of \gls{egs} market opportunity, see Figure \ref{fig:map_dh_demand_and_egs_suitability} for district heating demands and \gls{egs} suitability, and Figure \ref{fig:map_capacity_factors} for solar and wind capacity factors.

The lens of cost reductions highlights the technological progress needed for \gls{egs} viability. 
Figure \ref{fig:map_capex} shows a more system-centric perspective by showing the drilling \gls{capex} at which \gls{egs} becomes viable. 
While this perspective aligns largely for electricity-, noticeable deviations appear for heat (co-)generating \gls{egs}. 
In Sweden, Finland, and the Baltics, \gls{egs} capacity is built at very high drilling costs of 4,500-5,500$\,$\euro/kW. 
Conversely, more Central European regions require drilling costs to decrease to 2,000-4,000$\,$\euro/kW. 
Only minor cost reductions are required to achieve both of these cost levels. 
However, the difference in \gls{capex} reveals that \gls{egs} adoption in Scandinavia and the Baltics is driven not by geological suitability but by the inefficiency of alternative low-carbon supply.

\subsection*{\textbf{Higher Renewable Costs and Larger Roll-Out of District Heating Maximize Market Opportunity}}
\label{subsec:sensitivity}

\begin{figure*}[t]
    \centering
    \includegraphics[width=\textwidth]{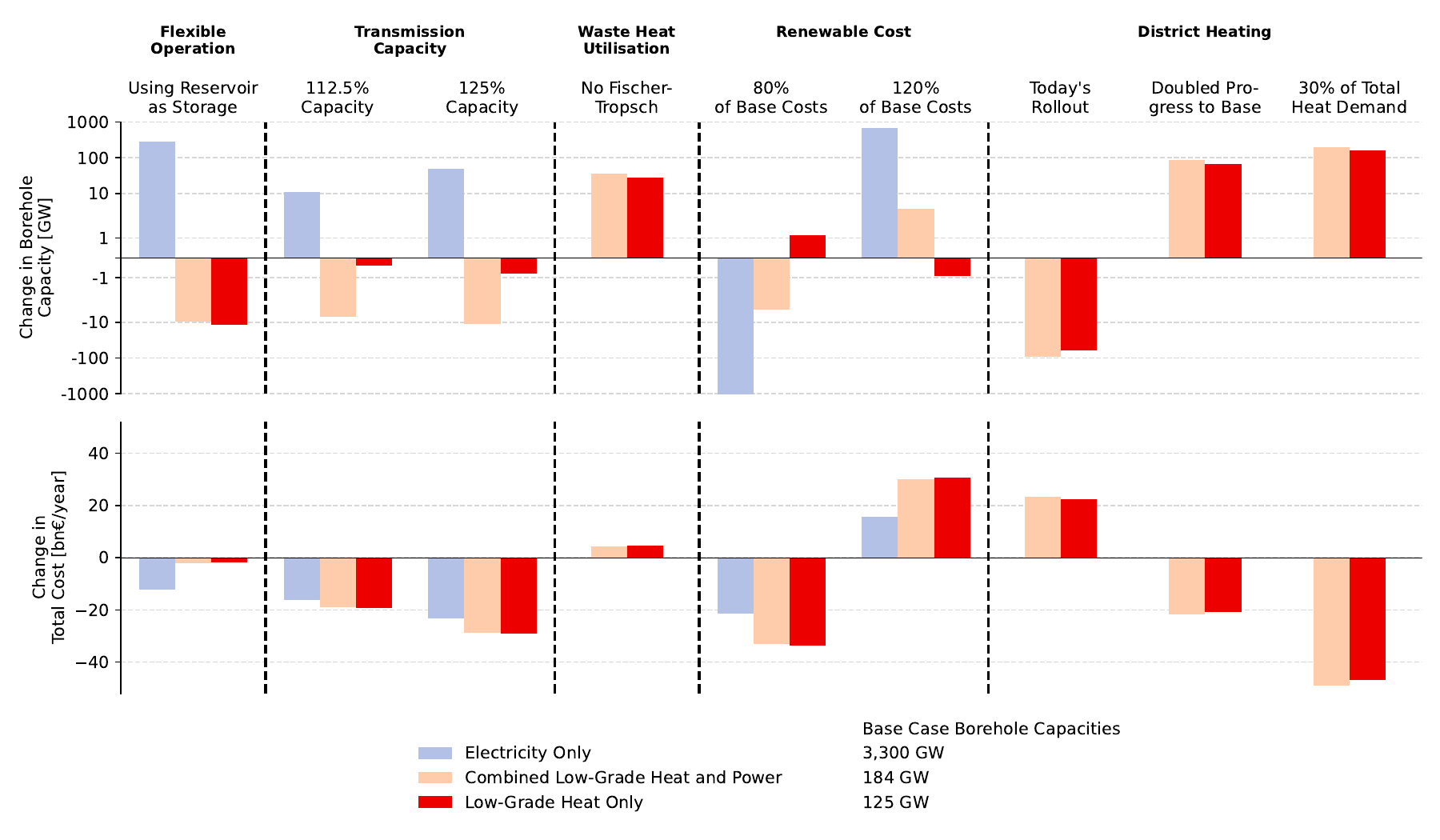}
    \caption{The change in borehole capacity and total system cost is evaluated for given sensitivities relative to \gls{egs} base cases, which include \gls{egs} under the same system assumptions as outlined in the previous sections.
    The base cases for heat-generating \gls{egs} assume a cost level of 40\% relative to 2020, while electricity-generating \gls{egs} is set at 20\%.}
    \label{fig:sensitivity}
\end{figure*}

In the following, we consider three sensitivities: 
Based on the previous analysis with high correlation of \gls{egs} adoption with electricity prices and heating demand, we \textit{(i)} test increased and decreased costs of renewable generation by 20\%, \textit{(ii)} vary the district heating rollout, and \textit{(iii)} disable the system's capacity to utilise waste heat for district heating. 
We further investigate the impact of \textit{(iv)} enabling flexible operation for \gls{egs} and \textit{(v)} the role played by transmission network capacity expansion.

We find the volume of district heating demand to be the most important factor in the rollout of heat-producing \gls{egs}. 
If district heating demand remains around current 2020 levels, rather than experiencing the moderate increases assumed in the base case, the market opportunity of \gls{egs} roughly halves, significantly hindering early-stage \gls{egs} technology learning. 
Meanwhile, the market opportunity of electricity generating \gls{egs} is more directly enhanced by any factor that improves its competitiveness with wind and solar. 
The same is true for \textit{flexible \gls{egs}}, where the geothermal reservoir is treated as storage, allowing the \gls{egs} plant to accommodate renewable intermittency better and thereby improve its market prospects. 
A more detailed overview of all tested sensitivities is shown in Table \ref{tab:sensitivities}.

Quantitatively, we find that for electricity-generating \gls{egs}, both flexible operation and a 20\% increase in renewable costs increase cost-optimal borehole capacity by approximately 1,000$\,$GW$_\text{th}$, compared to a \textit{base} capacity of roughly 3,300 GW$_\text{th}$ in the base case (Figure \ref{fig:sensitivity}). 
A 20\% decrease in renewable costs results in nearly matched reduction in borehole capacity.
This high sensitivity stems from the often minimal cost-advantage that electricity generating \gls{egs} has over other renewables. 
In contrast, under the same variation of renewable cost, the market opportunity for heat-producing \gls{egs} changes by less than 5\%, emphasising its substantial cost advantage over the next-best generation technology. 

The utility of flexible operation has already been shown in the United States to substantially improve \gls{egs} market opportunities \cite{ricks2024role}, and a similar dynamic is evident here. 
Flexible \gls{egs} can operate at a lower partial load during periods of high intermittent generation and then, after pressure and temperature in the reservoir have increased, generate beyond the plant's nameplate capacity. 
While potentially advantageous, increasing reservoir pressure may elevate the likelihood of earthquakes \cite{deichmann2009earthquakes}. 
Therefore, the future feasibility of this approach remains uncertain, though flexible \gls{egs} is still explored in research as a method for improving market viability. 
With flexibility, we observe an increase of around 300 GW$_\text{th}$ (around 10\%) in borehole capacity, which is lower than the average 30\% increase reported in \cite{ricks2024role}. 
This difference could be attributed to various factors, including more pessimistic assumptions about reservoir performance in this analysis or differences in geological or other renewable geographic suitability. 
For heat-generating \gls{egs}, flexible operation decreases borehole capacity. 
This is because \gls{egs} capacity has already saturated the limited district heating demand, and the system uses flexible operation to provide the same energy volume at higher capacity factors.
For broader context, the effect of flexible operation on \gls{egs} adoption is shown for all evaluated levels of reductions. 
While these effects are meaningful, they are small relative to the overall magnitude of both \gls{egs} capacity and cost reduction (Figure \ref{fig:flex_main}).

Heat-producing \gls{egs} is most affected by the rollout of district heating and the removal of Fischer-Tropsch process waste heat as a competing source of low-grade heat. 
This work uses a \textit{base case}, in which 30\% of the total potential to increase district heating demand to meet 60\% of all urban heat demand is met (see Methods Subsection \ref{subsec:energy_system_model}).
It turns out that switching from today's district heating rollout to a highly optimistic scenario increases cost-optimal borehole capacity by around 200 GW$_\text{th}$. 
In this optimistic scenario, it is assumed that 60\% of all urban residential heat demand is met through district heating networks. 
In many regions, \gls{egs} is significantly more cost-efficient than alternative carbon-free heat supplies, making it unsurprising that further demand increases its market opportunity. 
However, the direct and almost linear relationship between district heating rollout and demand for heat-based \gls{egs} is noteworthy, as it directly influences the volume of scarce ground available for technology learning towards electricity generating \gls{egs}.

\section*{Discussion}
\label{sec:discussion}

\paragraph{\textbf{Two phases of \gls{egs} deployment}}
Our results quantify how cost reductions in \gls{egs} influence its market potential, both in terms of overall volume and the sectors it can supply.
These cost are expressed relative to 2020 levels estimated in \cite{aghahosseini2020hot}.

We find that when drilling costs reduce by less than 60\%, heat (co-)generation is crucial for \gls{egs} to be competitive. 
However, at a level of 40\% of current costs, a \enquote{tipping point} is reached, after which electricity-generating \gls{egs} can compete directly with wind and solar generation, unlocking an exponentially larger market opportunity. 
We refer to the periods before and after this tipping point as the first and second phases of \gls{egs} deployment.

The first phase is crucial with regard to \gls{egs} technology learning. 
Our results show that district heating demands enable around 100 GW$_\text{th}$ of \gls{egs} capacity expansion that is already cost-competitive with alternative supply in a carbon-neutral system. 
As such, this capacity requires only minor financial aid, as subsidies would primarily be needed to account for its carbon neutrality. 
From the relationship found between cost reduction and market opportunity, we assess that with a learning rate between 20 and 25\%, deployment during the first phase of \gls{egs} could suffice to unlock the second phase. 
This is slightly higher than the 20\% observed for solar generation \cite{nijsse2023momentum}, but it should, however, be considered in the context of this study's exclusive focus on Europe; 
\gls{egs} cost reductions unlock markets outside Europe, likely leading to less stringent demands on the learning rate.

In the second phase, electricity-generating \gls{egs} competes directly with wind and solar generation. 
Competition in electricity markets entails a significantly larger market opportunity, estimated at approximately 250 GW$_\text{th}$ when costs decrease to 40\% of current levels, and about 900 GW$_\text{th}$ when reduced to 20\%. 
During this phase, \gls{egs} gains access to a much larger market, which implies lower requirements on the learning rate relative to phase one, converging to approximately 20\%. 
Therefore, it appears likely that if \gls{egs} can build the momentum to reach phase two, that momentum could be sustained without additional public funding.

\paragraph{\textbf{Spatial Distribution of \gls{egs} Adoption}}
Our analysis highlights the stark differences in the spatial distribution of market opportunities for heat- and electricity-generating \gls{egs}. 
However, we find that only a small number of factors classify regions into four distinct roles with regard to \gls{egs}' pathway to market.

First are potential early \gls{egs} adopters. 
Surprisingly, in this case, the primary driver for adoption is not geological suitability but rather the high cost of alternative sources of carbon-neutral energy for district heating schemes. 
As a result, Scandinavia and the Baltics could see substantial investment in \gls{egs} at current cost levels of 4500\euro/kW$_\text{th}$–\\5500\euro/kW$_\text{th}$. 
The second role includes the majority of heat-generating \gls{egs} adopters, such as France, Benelux, Southern Germany, and the Balkans. 
In these regions, a market opportunity of approximately 10–30 GW$_\text{th}$ is already accessible at current cost levels, increasing to about 100 GW$_\text{th}$ if \gls{egs} costs decrease by up to 40\%. 
At this stage, \gls{egs} outcompetes most alternative carbon-neutral district heating options, making the demand for district heating itself the primary predictor of \gls{egs} adoption. 
The third kind of region sees investment into electricity-gene-\\rating \gls{egs} once drilling costs have reduced by around 60\%. 
These regions, which include Central-Eastern Europe, mostly see \gls{egs} adoption due to high geological suitability. 
The remaining regions—namely Iberia, England, Northern Germany, Southern Italy, and Western Po-\\land—are either geologically unsuitable, rich in renewables resulting in low energy costs, or lack district heating infrastructure. 
As a result, they provide limited opportunities for \gls{egs} adoption at any evaluated level of cost reduction.

In summary, we anticipate a geographically widespread uptake for early-stage \gls{egs}. 
Given subsequent cost reductions, this could be followed by a later concentration in regions with high geological suitability around Central Europe via electricity generating \gls{egs}. 
We also observe that the spatial distribution of existing \gls{egs} projects \cite{eyerer2020advanced} leaves many of the market opportunities identified here untapped. 
While some of this potential should be realised as carbon constraints increasingly reveal the identified distribution of \gls{egs} opportunity, policy interventions could help accelerate this process.

\paragraph{\textbf{Risks and Opportunities During \gls{egs} Rollout}}

Our analysis quantifies the sensitivities of \gls{egs} market opportunities and finds them to be primarily influenced by district heating rollout and the future cost of renewables. 
In decreasing order of importance, secondary factors include waste heat utilisation, flexible operation, and transmission grid reinforcement.

Regarding district heating, this work assumes a base scenario in which district heating will experience some limited rollout beyond current levels. 
For comparison, we also test a scenario where district heating remains at current levels and find that the market for high-cost \gls{egs} decreases drastically, from 200 GW$_\text{th}$ to around 100 GW$_\text{th}$. 
Moreover, the spatial distribution of this reduced demand further exacerbates the disparity between the spatial distribution of heat-producing and electricity-producing \gls{egs} market opportunities. 
This is because, without further rollout, current district heating demand is even more concentrated in Eastern European and Scandinavian countries, while the market opportunity for electricity generation remains in more central regions. 
Therefore, in the \gls{egs} context, district heating expansion not only provides a means to expand demand for \gls{egs}, but also serves as a vehicle to prevent regional lock-in effects.

\gls{egs} has a unique interaction with transmission capacity. 
When generating electricity only, it exhibits similar dynamics to other renewables: 
if transmission capacity expansion is allowed, the model partially pays for grid reinforcement through lower expenses in generation capacity, enabled by better geographic suitability \cite{brown2018synergies}. 
However, when \gls{egs} is used to meet heat demands, this interaction no longer applies. 
For generation of combined low-grade heat and power, enabling transmission expansion reduces its market opportunity. 
Due to the difficulty of transporting heat, \gls{egs} cannot relocate to take advantage of the spatial flexibility offered by larger transmission capacity. 
As such, the common association of transmission expansion benefiting renewable energy holds for \gls{egs} only to a very limited extent. 
If large cost reductions fail to materialise, transmission expansion should not be considered a catalyst for \gls{egs} deployment.

\section*{Conclusion}
\label{sec:conclusion}
This work outlines a potential trajectory for the large-scale adoption of \gls{egs} in a carbon-neutral, multi-sector energy system for Europe.

We show that heat (co-)generating \gls{egs} would be already cost-competitive with today's drilling costs in regions with district heating demand and low renewable resources. 
Under minor cost reductions, these conditions extend to large parts of Europe, including Central, Eastern, and Northern regions, and could serve as a market entry for \gls{egs} of around 100–200 GW$_\text{th}$ capacity. 
If drilling costs reduce further to around 40\% of current levels, \gls{egs} that generates only electricity becomes cost-competitive with other renewable resources, enabling large-scale adoption in Europe of 20–100 GW$_\text{el}$ power capacity. 
While larger in volume, we find the vast majority of this power capacity to be located in Central European regions where high geological suitability and low renewable resources overlap. 
During early-stage cost reduction, heat (co-)generating \gls{egs} serves as a `runway`, where, through technology learning, the necessary cost reductions for competition in electricity markets are achieved.

We identify several factors that either impede or facilitate this process. 
The spatial distribution of market opportunities for heat-generating \gls{egs} and electricity-only \gls{egs} varies widely, with the former concentrated in Eastern and Northern Europe, and the latter more centrally located.

We conclude that \gls{egs} is cost-competitive for heat generation in regions with highly suitable geology and district heating networks, while its role in electricity generation depends on future drilling cost reductions. 
If supported through policies promoting heat generation, \gls{egs} could benefit from technology learning, reducing costs and making electricity-only generation from \gls{egs} competitive. 
Such policies would further enable \gls{egs} to reduce reliance on storage and renewable build-outs, which often face public opposition.

\section*{Methods}
\label{sec:methods}
\subsection*{Energy System Model} \label{subsec:energy_system_model}

We conduct our analysis in the sector-coupled model of the European-level energy system PyPSA-Eur \cite{neumann2023potential}.
It optimises investment and operation of energy generation, transport, storage and conversion over a full year with a three-hourly resolution, covering demands from the power, heat, transport, industry and agriculture sectors while maintaining a neutral carbon balance.
We provide a brief overview of the central features but refer to \cite{neumann2023potential} for a more detailed discussion.

To reflect the energy system's geographical context in terms of renewable potentials, energy transport and demand, the model represents Europe in 72 zones.
Each zone is assigned a weather-based time series of renewable maximum generation output per installed capacity, based on the open-source Python tool \textit{atlite} \cite{hofmann2021atlite}.
For the transmission grid, we have opted, based on the small impact observed in our sensitivity analyses, to prohibit the model from expanding the line capacities in an effort to reduce computational overhead.

Crucially, the model features EGS-relevant power, district heating, and low-temperature industrial heat demands.
Electricity demand is increased relative to today's levels due to the electrification of other sectors, leading to a total demand of around 11PWh (Fig. \ref{fig:model_overview}).
Heating demand aggregates to a yearly total of 3PWh, only 0.6PWh of which is district heating, with the remaining demands being either rural (with insufficient population density for communal heating) or non-residential sectors.
The model can also increase the volume of heat supplied to district heating nodes and use the excess heat for direct air capture.
To estimate the magnitude and distribution of heat demands, the model employs the methods outlined in \cite{zeyen2021mitigating}.

Yearly total heating demand per region is disaggregated from country-level figures based on the \textit{Integrated Database of the European Energy System} (JRC-IDEES) \cite{jrcidees}, Eurostat \cite{eurostat}, complemented by additional figures for Switzerland \cite{eidgenoss} and Norway \cite{norway_stats}.
These aggregate numbers are then spatially distributed by population density.
Further, the NUTS3 dataset \cite{nuts3} is used to assign the share of the population living in urban areas per region.
For the demand classified as urban, district heating shares are assigned based on today's district heating shares at the country level \cite{euroheat}.
However, further district heating rollout is widely seen as an ingredient of a cost-efficient residential heat supply in Europe, and as such the model adds our assumptions of future district heating demand to these demands.
The degree of progress assumed is quantified through the share of urban residential district heat demand that is met by \gls{egs}.
Our default assumption is a progress of 30\% between today's rollout and the maximum plausible share of district heat.
This share is assumed to be 60\% of urban residential demand.
An overview of country-level district heating demand levels is given in Figure \ref{fig:district_heating_rollout}.
Additionally, the model considers low-temperature industrial heating demands, which are lumped together with district heating demands but, in magnitude, are much smaller.
Whenever the present text refers to \textit{district heat}, it includes this additional demand for low-temperature industry.
The incumbent technologies that supply these demands are biomass, gas boilers, CHPs, solar thermal or waste heat from fuel synthesis.
Additionally, we assume thermal storage to be available for demand shifting.

In terms of investment and operational costs, the model faces the conflict between the single-year optimisation horizon and the need to account for the value of technologies accrued over their lifetimes, which exceed a single year many times over \cite{neumann2023potential}.
Therefore, the model is provided with annuitised yearly payments using the annuity factor
\begin{equation} \label{eq:annuity}
    \alpha = \frac{1 - (1 + \tau)^{-n}}{\tau}
\end{equation}
where $n$ is an asset's lifetime and $\tau$ is the discount rate, assumed to be 7\% throughout the study.
In total, capacity expansion is costed by
\begin{equation} \label{eq:capital_cost}
    (\alpha + \text{FOM}) \cdot \text{CAPEX}
\end{equation}
where FOM is the Fixed Operation and Maintenance Cost [\%] and CAPEX is the capital expenditure [\euro/kW].

\subsection*{Technology Learning Rate} \label{subsec:tech_learning}

We use a standard formulation of technology learning, where a learning rate quantifies the technology cost reduction per doubling of installed capacity of that technology \cite{basalla1988evolution}.
Starting with the \textit{power law} equation
\begin{equation} \label{eq:power_law}
    C(x) = C_0 \cdot x^{-b}
\end{equation}
where $C(x)$ refers to the cost at cumulative installed capacity $x$, $C_0$ to the initial costs, and $b$ to a learning exponent, we obtain the learning rate from 
\begin{equation} \label{eq:learning_rate}
    LR = 1 - 2^{-b} \: .
\end{equation}
It is this learning rate that is referred to in the text.
In the following, the \textit{cost reduction factor} refers to
\begin{equation}
    \text{cost reduction factor} = \frac{cost_{\text{original}}}{cost_{\text{reduced}}} \: ,
\end{equation}
i.e. a cost reduction factor of 2 refers to halved costs.
For the initial capacity $C_0$, we use the capacity installed by the model without cost reductions, i.e., at cost reduction factor 1 \cite{geofactsheet}.
From this, at each level of cost reduction $y$, we obtain the respective learning rate using the respective installed capacity by the model $C_y$, such that
\begin{equation} \label{eq:our_learning_rate}
    b = \frac{\text{ln}(1/y)}{\text{ln}(C_y / C_0)} \: ,
\end{equation}
and derive the learning rate $LR$ from Equation \ref{eq:learning_rate}.
Note that $y$ is larger than 1, which we accommodate by making it the denominator.

\subsection*{Modelling Enhanced Geothermal Systems} \label{subsec:modelling_egs}

We include \gls{egs} in the energy system model in three different modes each with different energy outputs - electricity, combined low-grade heat and power or low-grade heat only.
All three share the same subsurface modelling, but differ in the surface plant's configuration. 

Considering the surface plant first, for electricity generation, we simulate an organic Rankine cycle (\gls{orc}), which is a standard choice for electricity generation at low temperatures, utilising the lower boiling point of the organic fluid relative to water \cite{tester2006future, ricks2024role, aghahosseini2020hot}.
For the efficiencies of the ORCs, we use estimates based on regression of existing plants to be around 7–8\% at approximately 105$^\circ$C and plateau at approximately 15\% from 200$^\circ$C onward, which could further increase through modifications to the plant \cite{tester2006future, ahmadi2020applications}.
Our model works with a default value of 12\% efficiency, aligning with temperature assumptions in the underlying geothermal potential data (see Subsection \textit{Data on Geological Suitability}).
Later in this subsection, we discuss the implications of our results for \gls{egs} operating at other efficiencies.
While ORCs are a mature technology, uncertainty in terms of realistic cost assumptions remains.
Various sources assess plant costs to be between 1300\euro/kW$_{\text{el}}$ and 2000\euro/kW$_{\text{el}}$ \cite{tartiere2017world, frey2023techno, ricks2022value, aghahosseini2020hot}.
In our model, we use \gls{egs} capital expenditure of 1500\euro/kW$_{\text{el}}$, which aims to reflect an average of these observed values paired with modest technology learning.
The \gls{orc} is also assumed to have an average technical availability that allows it to deliver, on average, 93\% of nameplate capacity \cite{ricks2024role}.
In general, we assume that the \gls{orc}'s ramp rates are large relative to the temporal resolution of the model, effectively imposing no constraints on dispatch between time steps \cite{benato2015analysis}.

For combined low-grade heat and power operation, this setup is enhanced according to \cite{eyerer2020advanced}.
In this case, the model can modulate the working fluid flow between an electricity-generating cycle (ORC) and a heat-exchanger cycle, which transfers heat content to meet the required demand temperatures.
In this setting, \cite{eyerer2020advanced} report a broad operational range down to 15.3\% partial load.
Based on this, we assume that our modelled absence of partial load constraints introduces only a minor error.
Using an estimation made in \cite{frey2023techno}, we cost the additional infrastructure to supply district heat at 25\% of the \gls{orc} cost.
While this value, i.e., the cost of the heat integration, can vary due to a number of factors, we assume a default value here to avoid cluttering the analysis.
PyPSA-Eur, without \gls{egs}, employs a 15\% loss of district heat during distribution.
Our model concatenates this with an \gls{egs} heat generation loss of 5\%.
During capacity expansion, we assume a fixed ratio between heat and electricity capacity generation. 
The fixed ratio is given by their efficiencies $\eta_{\text{dh}} / \eta_{\text{el}} \approx 8$, with the two individual values taken from \cite{tester2006future, frey2023techno}.
However, we do not constrain the ratio between heat and electricity generation during dispatch.

Finally, for the generation of low-grade heat only, the ORC part is removed, leaving only the heat exchanger cycle operating at an assumed 95\% efficiency. 
Our modelling uniformly sets heat generation temperatures to 150$^\circ$C, which helps streamline our analysis but necessitates additional steps to generalise the results to other production temperatures.
At 150$^\circ$C working fluid temperature, which is a typical value in existing geothermal projects albeit low for projected future projects, the electricity generation efficiency is around 12\% \cite{tester2006future}.
The formulation of our capital cost (\euro/kW) allows us to remain agnostic to different temperatures.
However, this approach misses how different temperatures change the distribution of CAPEX between the geothermal well and the \gls{orc} plant, which can complicate the utilisation of our findings in real \gls{egs} projects.
To ensure interpretability, we recommend the following protocol to infer the implications of our findings for an arbitrary drilling CAPEX $c_x$ at an arbitrary temperature $x$.
Statements made in the paper are formulated through drilling CAPEX $c_{150^\circ\text{C}}$, expressed in units \euro/kW$_{\text{th}}$.
To interpret the respective findings in terms of the total plant cost, i.e., aggregating subsurface and surface parts, we calculate the overall CAPEX expressed in \euro/kW$_{\text{el}}$ as follows:
\begin{equation} \label{eq:total_plant_cost}
c_{\text{total}} = \frac{c_{150^\circ\text{C}}}{\eta_{150^\circ\text{C}}} + c_{\text{orc}} \: ,
\end{equation}
where $\eta_{150^\circ\text{C}}$ is the \gls{orc} plant's power generation efficiency at 150$^\circ$C, which is 12\%.
At a different temperature level, electricity is generated at efficiency $\eta_{x}$, and the respective total cost follows from the analogue of Equation \ref{eq:total_plant_cost}.
It follows that for a geothermal well generating heat at temperature $x$, the relevant CAPEX to compare against in our analysis is given by 
\begin{equation} \label{eq:plant_mapping}
c_{150^\circ\text{C}} = c_x \frac{\eta_{150^\circ\text{C}}}{\eta_{x}} \: .
\end{equation}
Note that this is only relevant for power-generating \gls{egs}, since low-grade heat usages convert incoming heat at extremely high efficiencies to the required output temperatures.

While the surface part differs, all three share the same simulation of borehole operation. 
The subsurface part of the \gls{egs} plant is split into the injection and production wells, with capital expenditure assigned equally between them. 
We take the perspective of the whole energy system, and as such, the geothermal well is treated as a firm generator of heat, assumed to be capable of modulating its energy output on timescales smaller than the model's temporal resolution of three hours. 
If flexible operation is activated, we additionally treat the geothermal reservoir as storage to represent the build-up of pressure and temperature when backpressure is applied to the production well \cite{ricks2024role}.
The physics of that process are not included in the model, but they are assumed to justify the modelled storage capacity.
In this, we assume the reservoir can accommodate a build-up of 24 hours, which can be discharged through a generation surplus of up to 25\%.
This temporal extent of flexible \gls{egs} is slightly larger than the temporal extent of flexible operation tested in the field \cite{kelkar2016lessons}, but smaller than in similar studies \cite{ricks2024role}.
Finally, we model the efficiency of the geothermal heat extraction cycle similarly to \cite{ricks2024role}, as a function of local deviations from the ambient temperature at the plant's design point temperature.
The magnitude of the changes in technical availability is informed by measurements taken at the DORA-1 geothermal plant in Turkey \cite{karadas2015multiple}.

Investment costs for \gls{egs} are processed as outlined in Equations (\ref{eq:annuity}) and (\ref{eq:capital_cost}). 
Marginal costs are omitted for \gls{egs} plants, as they are assumed to be small relative to the investment costs.

\subsection*{Data on Geological Suitability}

Suitability for \gls{egs} is highly heterogeneous across Europe. 
We use data from \cite{aghahosseini2020hot} to represent the spatial aspects of geological suitability in the model and refer to that paper for a more elaborate methodological description, but provide a brief overview here. 
Building on the protocol proposed in \cite{beardsmore2010protocol}, the 1$^\circ$$\times 1^\circ$ gridded data provided in \cite{aghahosseini2020hot} estimates, for each square, the cost-optimal \gls{egs} depth by optimising the trade-off between higher costs through deeper drilling and lower surface plant costs achieved by higher temperatures of geological resources.
Further, \cite{aghahosseini2020hot} makes assumptions about future cost reductions resulting from technology learning.
Based on the projected exponential increase in \gls{egs} capacity towards 2050, paired with a technology learning rate of 7.5\% at each doubling of built capacity, the paper asserts the cost reductions used in the present paper (see Figure \ref{fig:schematic_cost_reduction}, reaching a maximum cost reduction factor of 5 in 2050).
To make our results more intuitive, we decouple the cost reduction from the specific year assigned in \cite{aghahosseini2020hot} and instead express it as a factor.
While \cite{aghahosseini2020hot} is supplemented only by country-level data, courtesy of the author, we were able to access the underlying gridded data for our experiments.
Each region in our model was assigned an \gls{egs} potential based on the overlay with its geometry.

\subsection*{Limitations}

We believe our work is subject to three main limitations. 
These are the simplified nature of our \gls{egs} simulation and potential estimation, the approximations made in the assignment of \gls{egs} costs to different regions, and the uncertainty about realistic energy transition pathways.

With regard to the first limitation, the physics that govern heat generation from the geothermal well are neglected in this work to accommodate the computational overhead of the model. 
Every \gls{egs} project is subject to uncertainties regarding the actual reservoir performance once drilling is completed, and rectifying subpar performance may require expensive geological engineering. 
In both the \gls{egs} potential dataset provided by \cite{aghahosseini2020hot} and this work, this risk factor is not priced into the drilling cost, leading to a potential underestimation of real costs. 
These cost factors are omitted since they depend on geological characteristics, the analysis of which exceeds what can feasibly be addressed in this study. 
Moreover, the geology of \gls{egs} itself creates further uncertainty in terms of generation over the plant's lifetime, including sudden structural changes in the reservoir that can lead to capacity reductions \cite{kelkar2016lessons}.
We assess this limitation to be within the standard approximations of large-scale models, which neglect similar nuances for other generation technologies.
When comparing results from our work to real-world estimates, we urge the reader to keep this in mind and consider comparing this work against risk-adjusted real-world costs.

However, the data underpinning the estimations of suitability for \gls{egs} used here are also subject to uncertainties beyond those in analogous datasets for other technologies. 
This is mainly due to subsurface temperature data, which is often found to differ from pre-drilling estimates. 
This issue is highlighted, for instance, by two datasets of subsurface temperature in the United States that exhibit substantially varying spatial distributions of values \cite{ricks2024role, aljubran2024thermal}. 
Therefore, our results should be interpreted as preliminary indications based on current knowledge of \gls{egs} suitability but may deviate from the true \gls{egs} opportunity.

This work also neglects the gradual nature of transition pathways in the European energy system. 
The uncertainty regarding how that transition may manifest motivated us to instead opt for a model with the common denominator of carbon neutrality and let the model optimise the energy system's configuration. 
The assumption of carbon neutrality lends a competitive advantage to low-emission technologies like \gls{egs}, which makes our findings less relevant in the current system. 
Especially in the heating sector, identified here as crucial for early \gls{egs}, marginal prices are much lower than those seen by the model, which primarily assumes conventional heating. 
However, we argue that, given the energy system's trajectory towards carbon neutrality, locating the analysis within this framing is the most sensible option, as it allows the model to weigh technologies against one another in the context most likely to reflect future conditions. 
Further assumptions about the future system, such as technology rollout in fuel synthesis or the adoption of district heating, were qualified through sensitivity analyses (see Subsection \textit{Higher Renewable Cost and Larger Roll-Out of District Heating Maximize Market Opportunity}).

Further approximations had to be made when assigning estimated \gls{egs} costs to regions of the network. 
Subsurface suitability can have a higher spatial resolution than the network, requiring the grouping of different potentials into a single potential assigned to one region. 
In the present work, each region is assigned the average \gls{egs} cost of all patches per region, weighted by heat potential. 
This method can mix regions of extreme suitability with regions of lower suitability, potentially missing low-cost \gls{egs}. 
While this is imprecise, it is a reasonable option in light of the low spatial granularity of the underlying dataset, which has a resolution of $1^\circ \times 1^\circ$, complicating regional assignment to begin with.

\subsection*{Usage of Large-Language Models}

The authors of this work have used large language models to improve the grammatical correctness, word choices, and clarity of the text.

\section*{Acknowledgements}

We are grateful to Wilson Ricks, Benjamin Pfluger, Gioia Falcone and Fabian Neumann for fruitful discussions and helpful suggestions.
We are particularly grateful to Arman Aghahosseini for providing a customised version of data from his research, tailored specifically to our context.
The authors would like to acknowledge the financial support of EPSRC (Engineering and Physical Sciences Research Council) and project partners of the INTEGRATE (EPSRC reference: EP/T023112/1) and DISPATCH (EPSRC reference: EP/V042955/1) projects.
For the purpose of open access, the authors have applied a Creative Commons Attribution (CC BY) licence to any Author Accepted Manuscript version arising from this submission.



% ...

\section*{Data and Code Availability}
\label{sec:code}

All code and data used is fully available at \\ \href{https://github.com/LukasFrankenQ/egs_market_integration}{https://github.com/LukasFrankenQ/egs\_market\_integration}.



\printglossary[type=\acronymtype]

\addcontentsline{toc}{section}{References}
\renewcommand{\ttdefault}{\sfdefault}
\bibliography{paper}

\appendix
\label{sec:appendix}
\renewcommand{\thefigure}{A.\arabic{figure}}
\renewcommand{\thetable}{A.\arabic{table}}

\begin{figure*}[t]
    \centering
    \includegraphics[width=\textwidth]{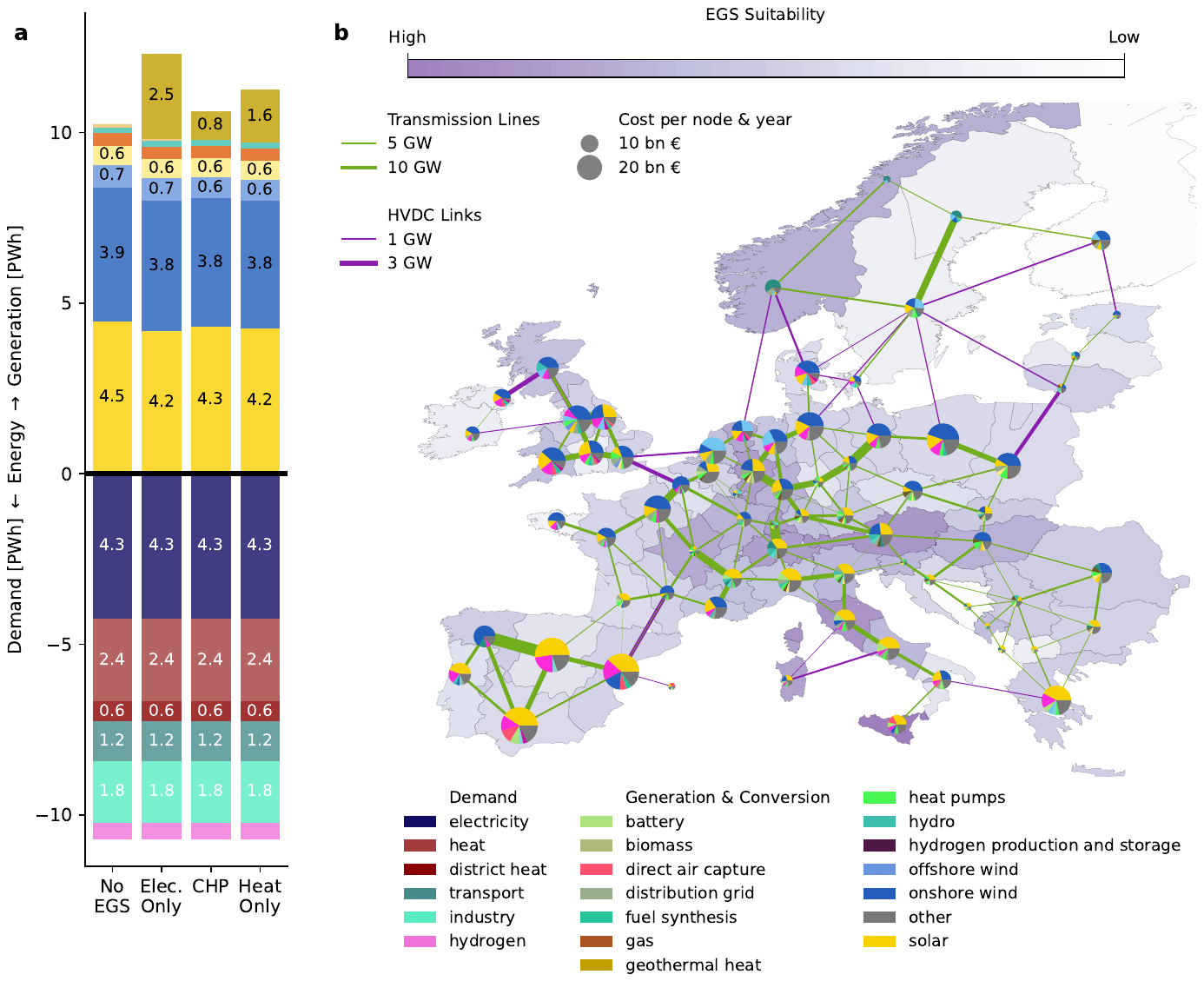}
    \caption{\textbf{a} Energy demand and generation for model runs without EGS, and with EGS generating $(i)$ only electricity, $(ii)$ combined low-grade heat and power and $(iii)$ only low-grade heat. \textbf{b} Annual cost of energy generation, conversion and storage for a model without EGS for each node. Note that for electricity only generating \gls{egs}, the volume of geothermal heat looks exaggerated as the low efficiency of power generation is yet to be applied. \gls{egs} suitability is taken from \cite{aghahosseini2020hot} and mapped to the model geometries.}
    \label{fig:model_overview}
\end{figure*}

\begin{figure}[t]
    \centering
    \includegraphics[width=0.95\columnwidth]{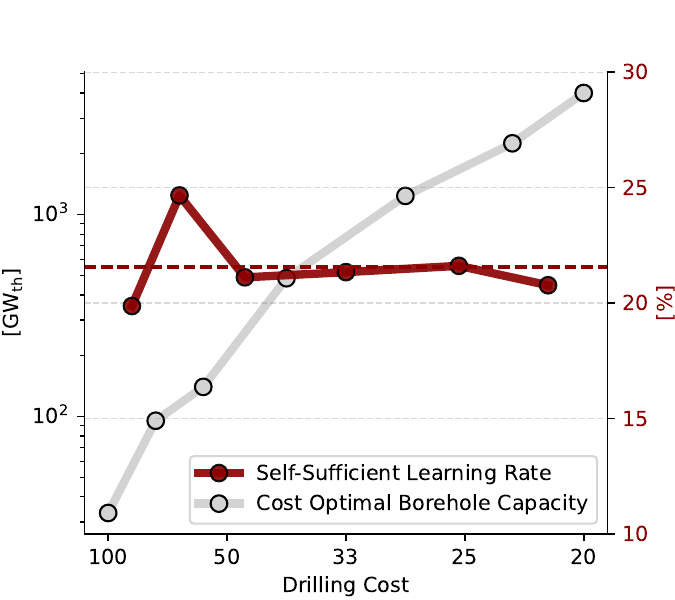}
    \caption{Cost-optimal borehole heat capacity summed over \gls{egs} generating electricity only and combined low-grade heat and power at different levels of cost reduction and learning rate required for \gls{egs} to itself to provide the opportunity for sufficient technology learning.}
    \label{fig:learning_rate}
\end{figure}

\begin{figure*}[t]
    \centering
    \includegraphics[width=0.95\textwidth]{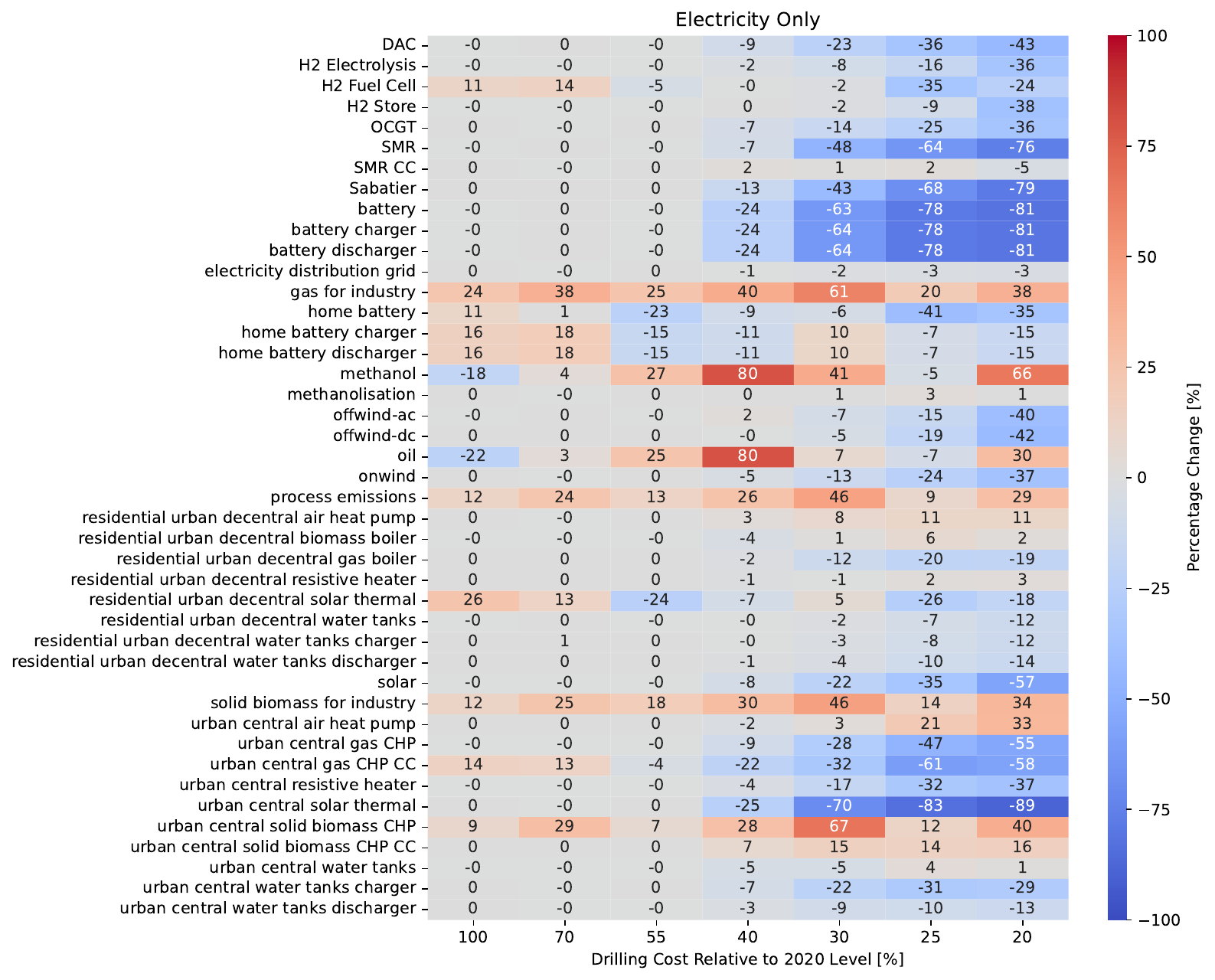}
    \caption{Percentages of changes in installed capacities of non-\gls{egs} technologies when electricity-only generating \gls{egs} is added at different levels of cost reduction.}
    \label{fig:percentage-displacement-elec}
\end{figure*}

\begin{figure*}[t]
    \centering
    \includegraphics[width=0.95\textwidth]{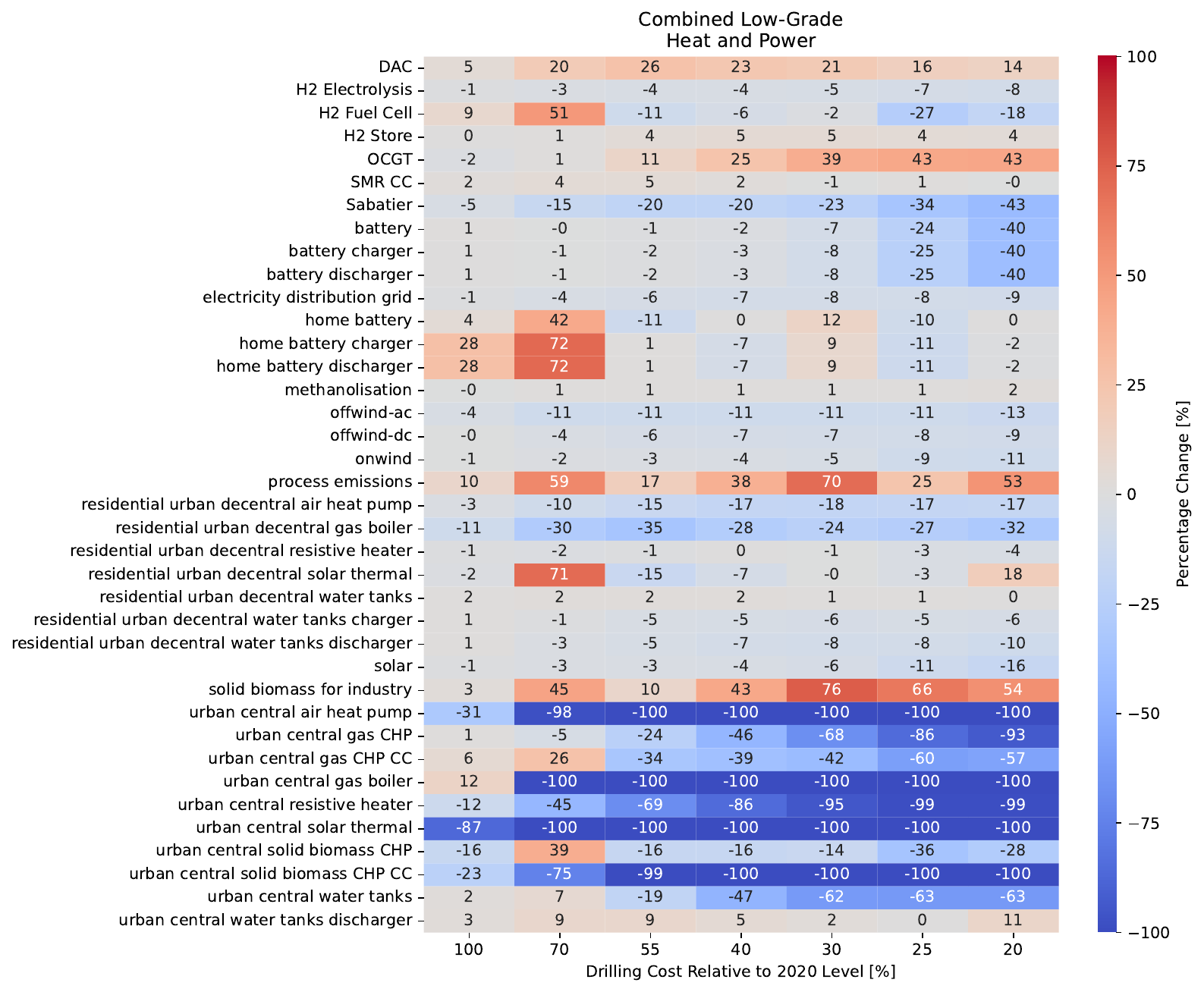}
    \caption{Percentages of changes in installed capacities of non-\gls{egs} technologies when \textit{Combined Low-Grade Heat and Power} \gls{egs} is added at different levels of cost reduction.}
    \label{fig:percentage-displacement-chp}
\end{figure*}

\begin{figure*}[t]
    \centering
    \includegraphics[width=0.95\textwidth]{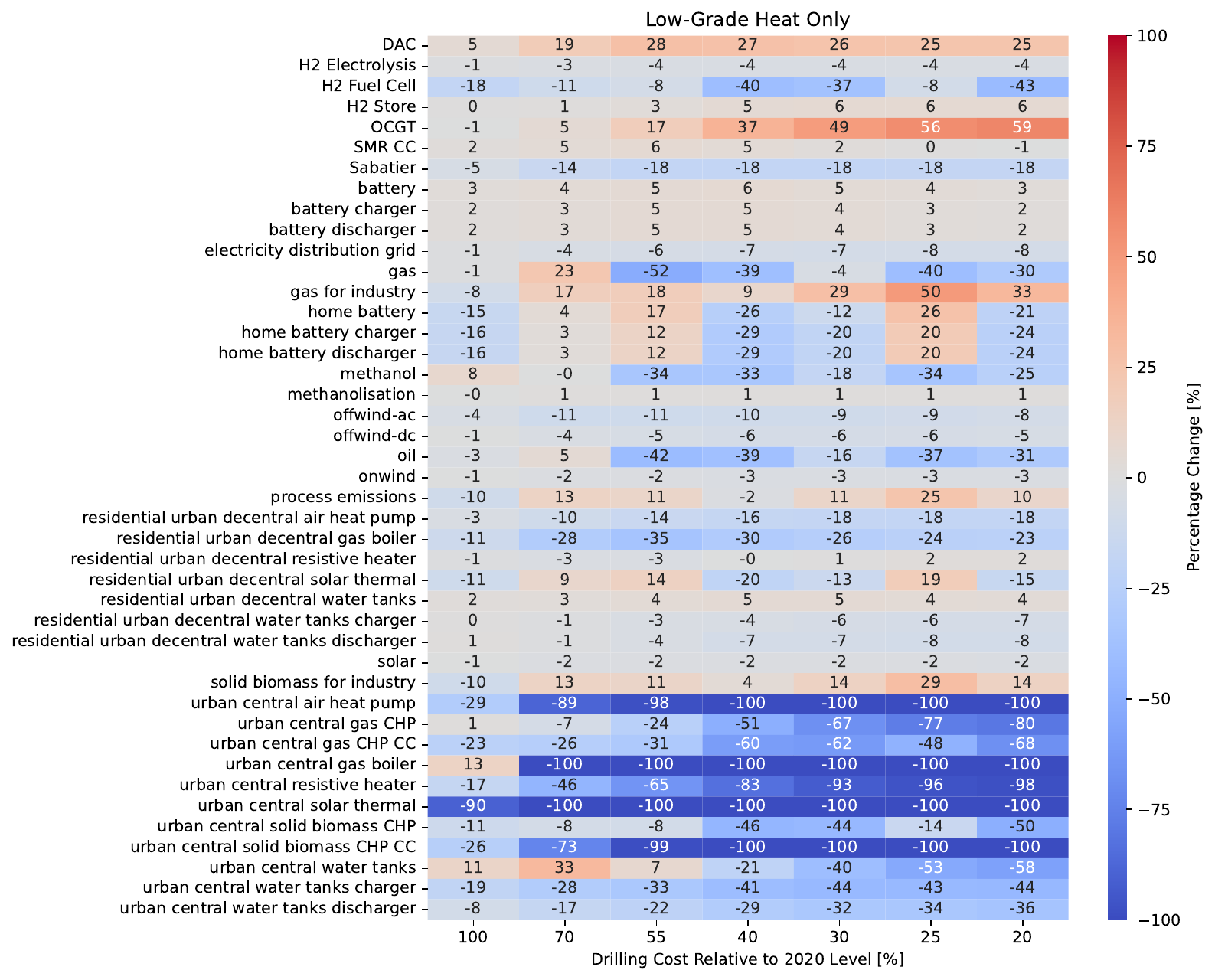}
    \caption{Percentages of changes in installed capacities of non-\gls{egs} technologies when \textit{Low Grade Heat} \gls{egs} is added at different levels of cost reduction.}
    \label{fig:percentage-displacement-dh}
\end{figure*}

\begin{figure*}[t]
    \centering
    \includegraphics[width=\textwidth]{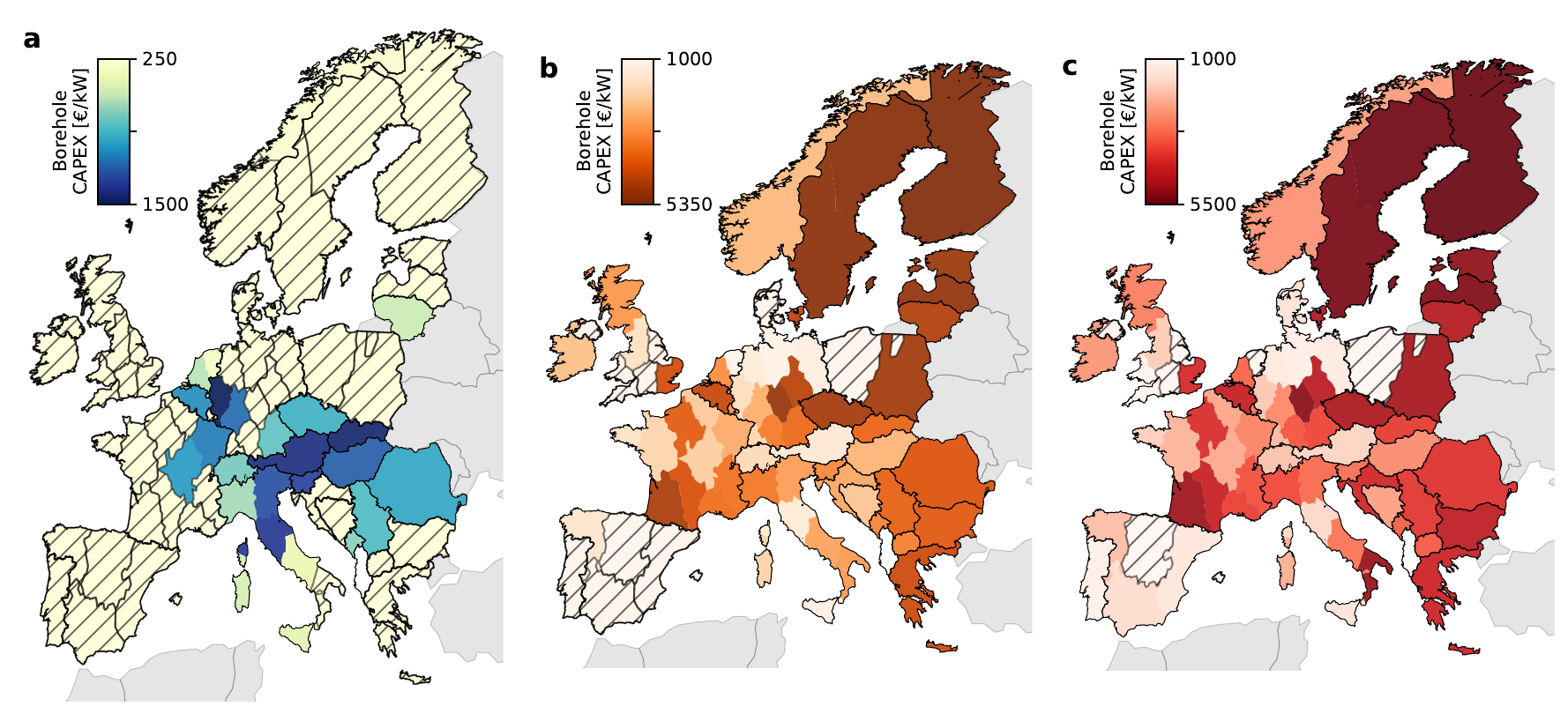}
    \caption{Borehole \gls{capex} at which \gls{egs} contributes 10\% of the respective demand per region. \textbf{a} for \textit{electricity only}, \textbf{b} for \textit{combined low-grade heat and power} and \textbf{c} for \textit{low-grade heat only}. Note the different spans of the respective colourmaps. The hashed region saw no EGS deployment in any modelled cost reduction factor.}
    \label{fig:map_capex}
\end{figure*}

\begin{figure*}[t]
    \centering
    \includegraphics[width=0.5\textwidth]{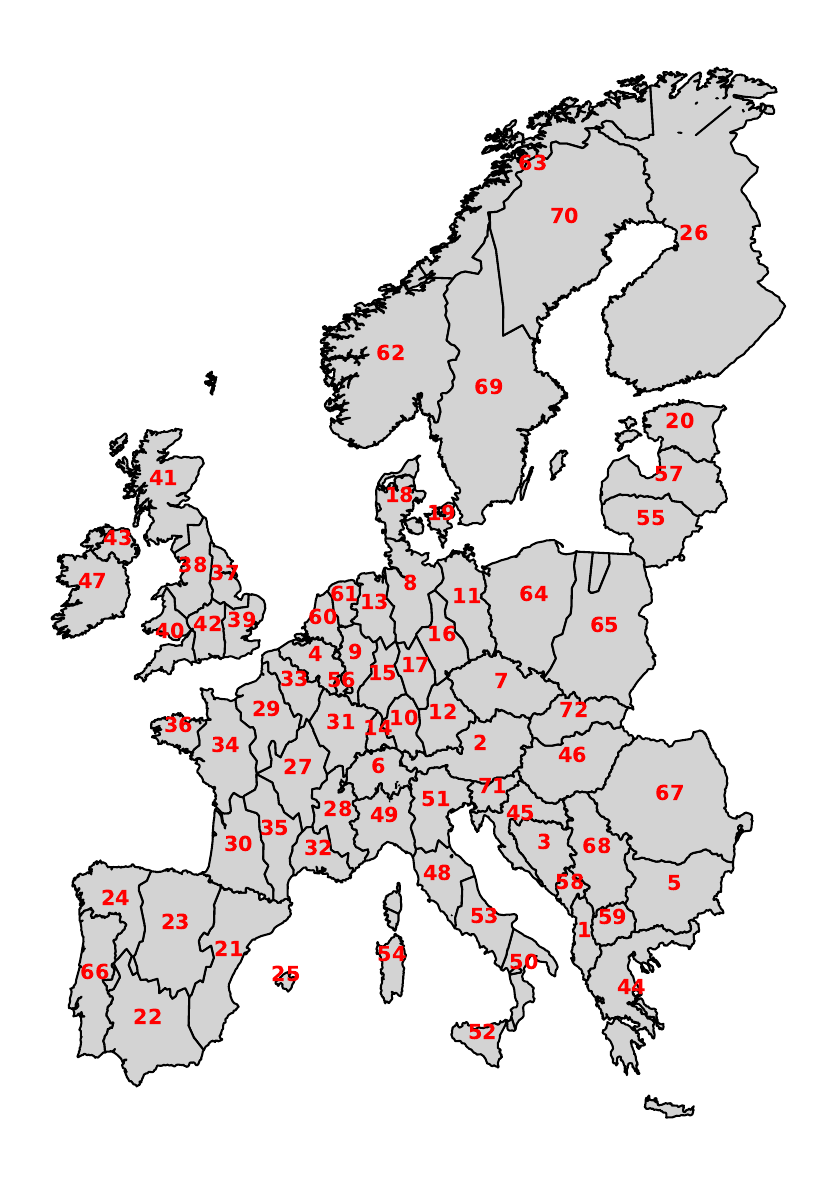}
    \caption{All regions modelled in the network with mock-names.}
    \label{fig:numbered_regions}
\end{figure*}

\begin{figure*}[t]
    \centering
    \includegraphics[width=\textwidth]{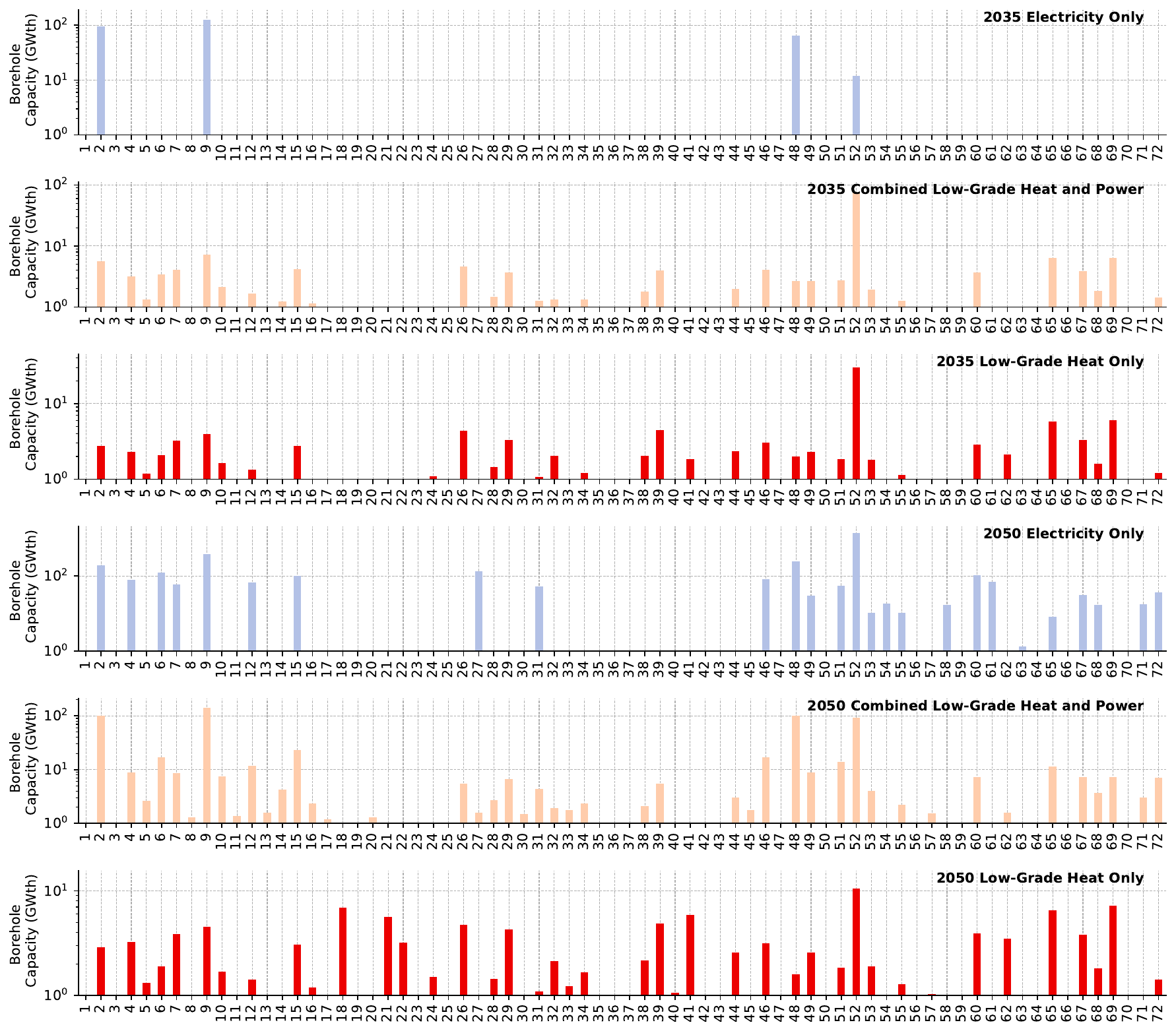}
    \caption{Injection well borehole capacities for all individual network regions. Labelling of regions as in Figure \ref{fig:numbered_regions}.}
    \label{fig:all_borehole_capacities}
\end{figure*}

\begin{figure*}[t]
    \centering
    \includegraphics[width=0.66\textwidth]{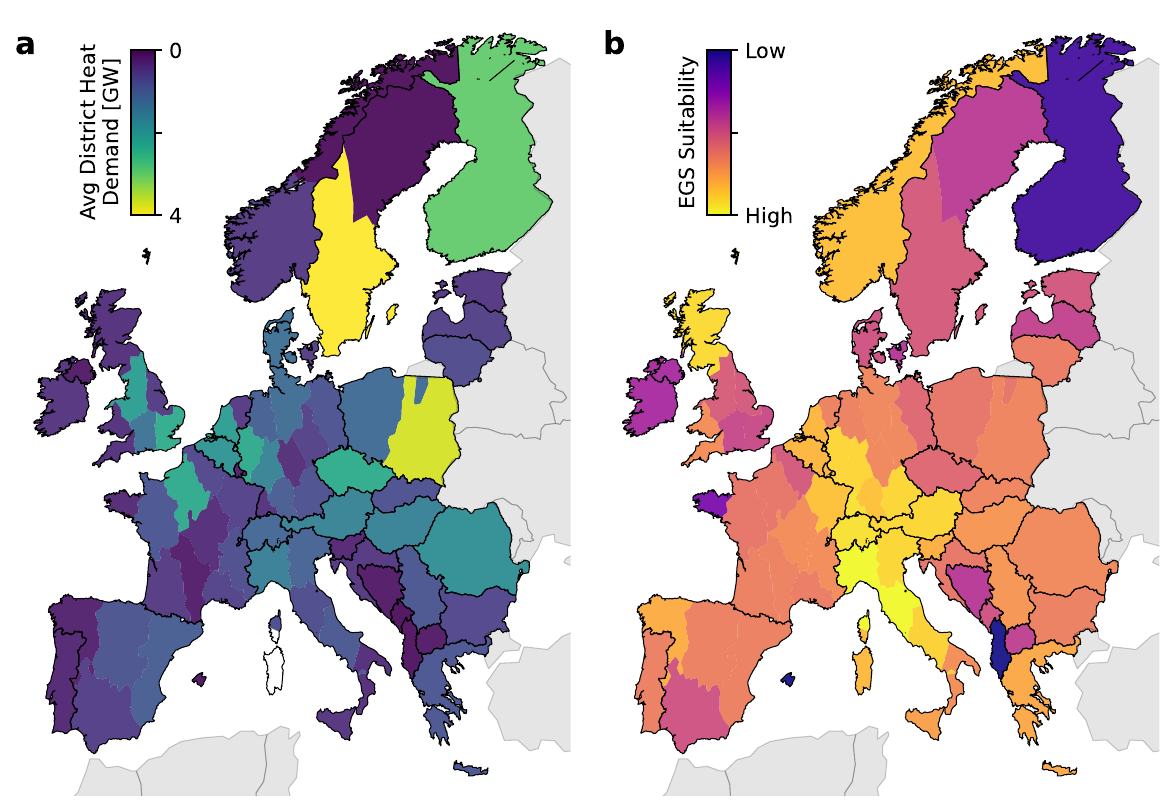}
    \caption{\textbf{a} Average district heating demand per region assuming a district heating rollout progress of 30\% (see Fig. \ref{fig:district_heating_rollout}) as assumed in the bulk of experiments and \textbf{b} relative EGS suitability .}
    \label{fig:map_dh_demand_and_egs_suitability}
\end{figure*}

\begin{figure*}[t]
    \centering
    \includegraphics[width=\textwidth]{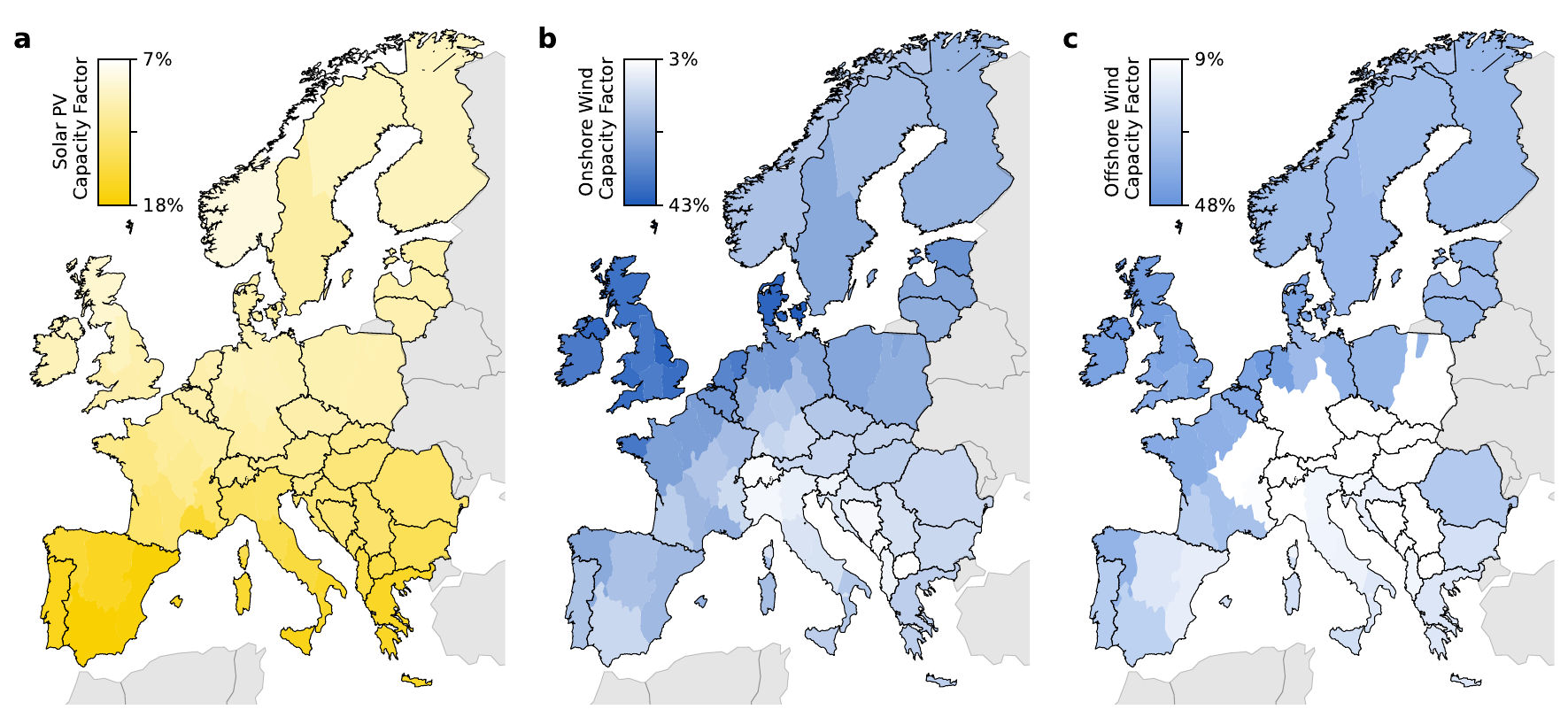}
    \caption{Yearly average capacity factors of \textbf{a} solar PV, \textbf{b} onshore wind and \textbf{c} offshore wind.}
    \label{fig:map_capacity_factors}
\end{figure*}

\begin{figure*}[t]
    \centering
    \includegraphics[width=\textwidth]{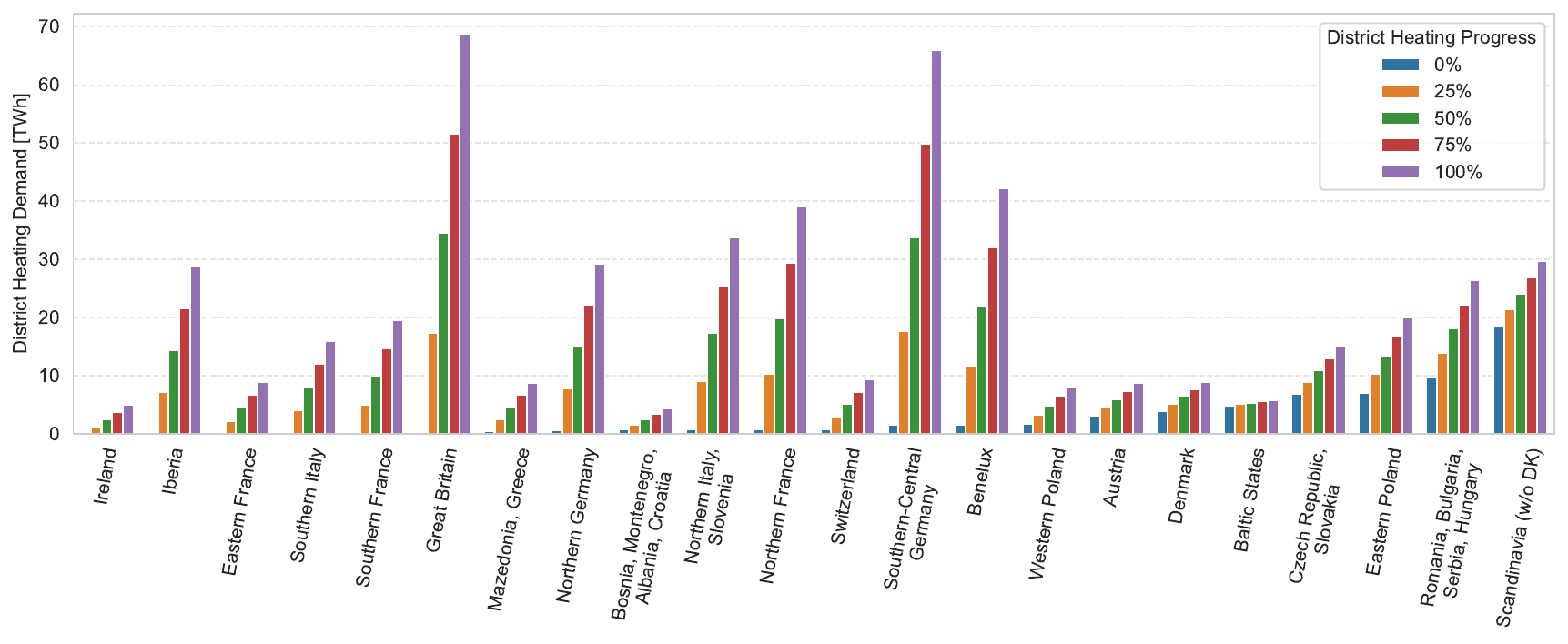}
    \caption{Total district heating demand per aggregated regions for different assumptions about district heating rollout.}
    \label{fig:district_heating_rollout}
\end{figure*}

\begin{figure*}[t]
    \centering
    \includegraphics[width=0.7\textwidth]{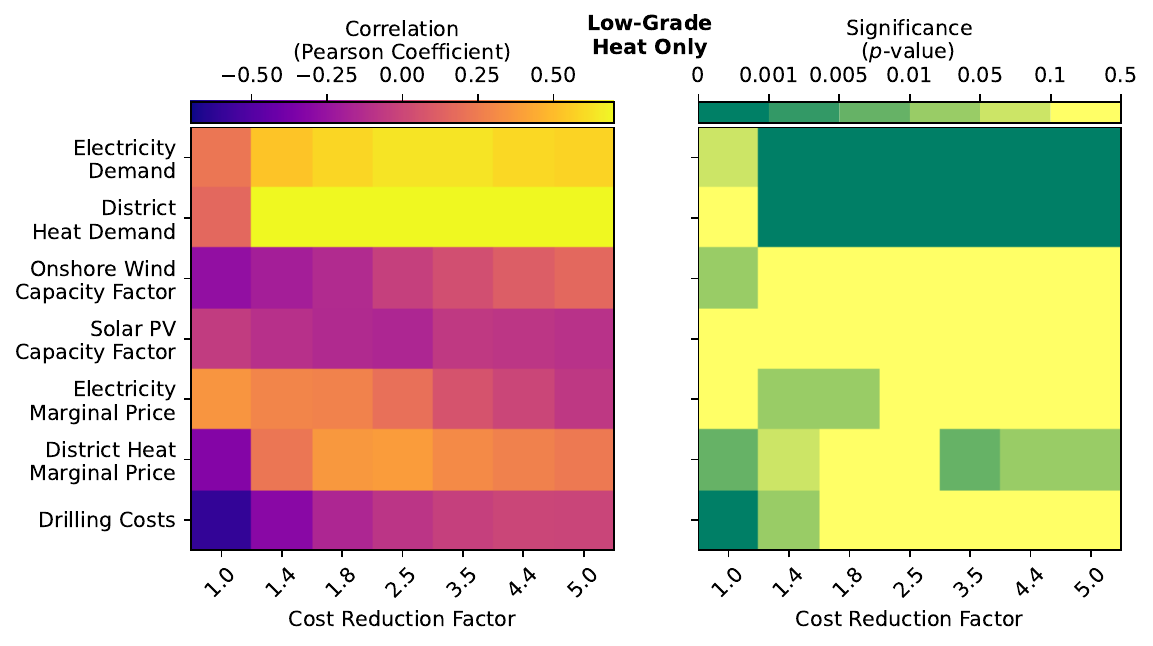}
    \caption{...}
    \label{fig:correlation_dh}
\end{figure*}

\begin{figure}[t]
    \centering
    \includegraphics[width=0.95\columnwidth]{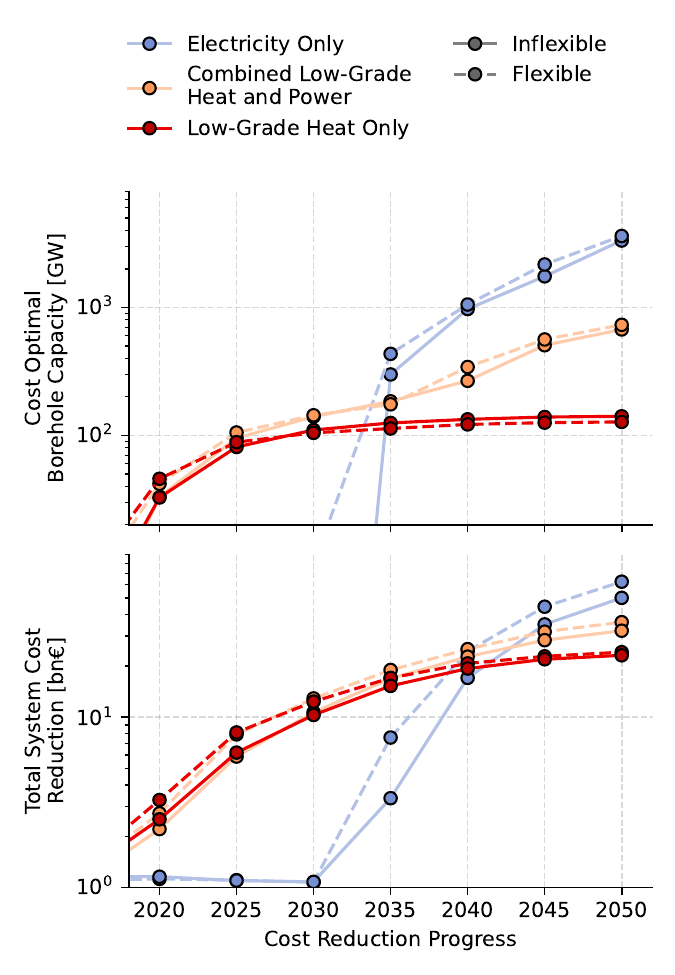}
    \caption{Built borehole capacity and total system cost change when \gls{egs} is operated flexibly.}
    \label{fig:flex_main}
\end{figure}

\begin{figure*}[t]
    \centering
    \includegraphics[width=\textwidth]{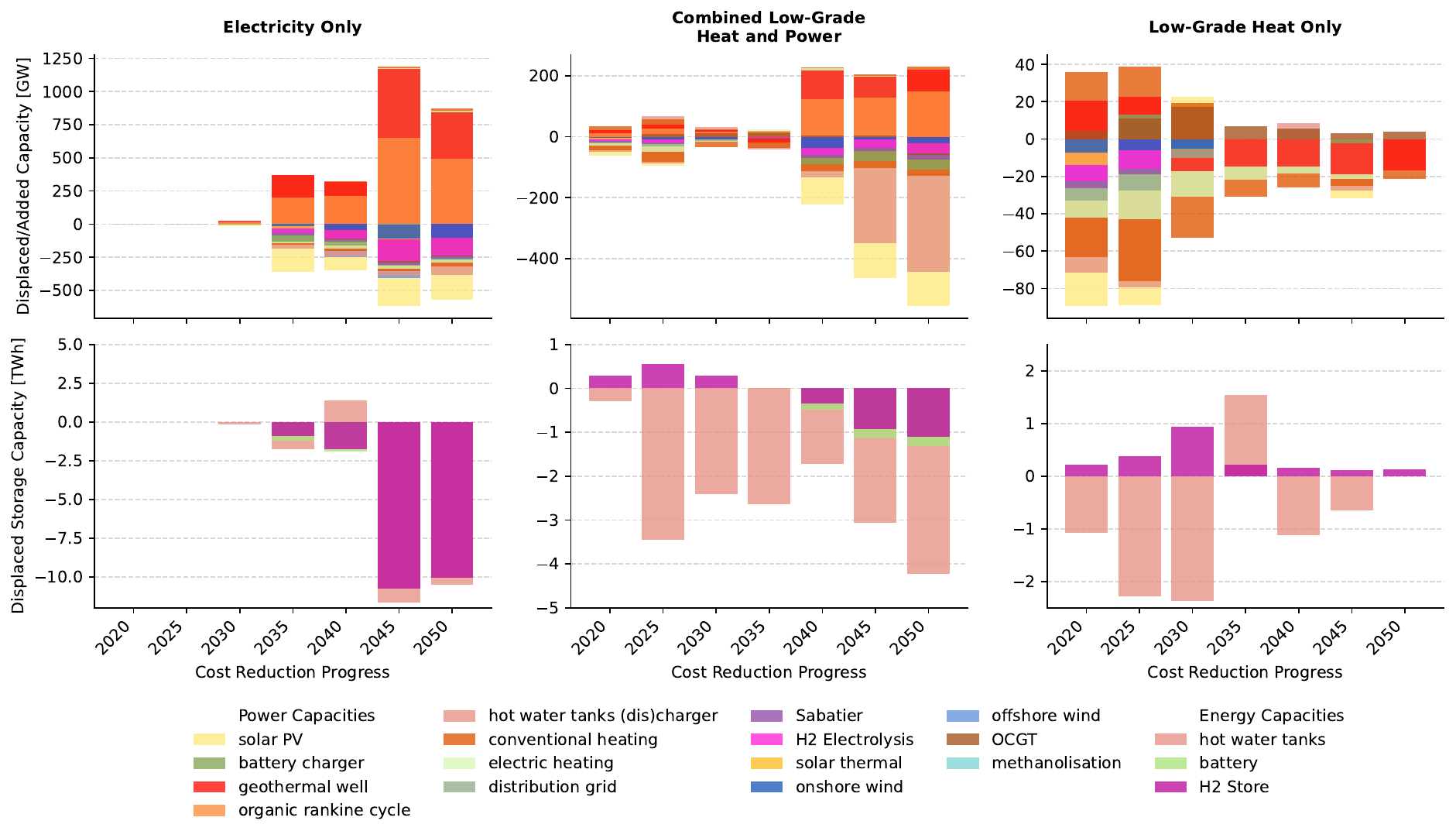}
    \caption{Changes of installed capacities by the model when flexible operation is switched on for different \gls{egs} cost reduction levels. The \textit{top} row refers to power capacities, the \textit{bottom} row to energy capacities. Positive values are added, negative ones are subtracted relative to a counterfactual where \gls{egs} is not available.}
    \label{fig:flex_displacement}
\end{figure*}

\begin{table*}[t]
    \centering
    \footnotesize
    \caption{Sensitivities tested relative to a \enquote{base case} of assumptions on the trajectory of the European energy system.}
    \vspace{0.3cm}
    \begin{tabular}{lcc}
        
        Sensitivity & Description & Tested Values \\ \hline \hline \\

        \multirow{6}{*}{Flexible Operation} & \multirow{5}{*}{\parbox{8cm}{The application of backpressure to the production well induces an increase in temperature and pressure within the geothermal reservoir, leading to enhanced power generation upon resuming production \cite{ricks2024role}. This approach, known as flexible EGS, allows EGS to synergise with variable renewable energy generation, thereby overcoming the constraint of operating solely as a base-load generator. In this study, we examine the effects of EGS operation after a period of reduced generation. }} & \multirow{5}{*}{\parbox{5 cm}{We incorporate a storage capacity into the reservoir, which can be charged for up to 24 hours. The discharge from storage is constrained to 25\% of the borehole's capacity.
        This represents an optimistic scenario, given the heightened risk of induced seismicity associated with such pressure increases \cite{deichmann2009earthquakes}.}} \\ \\ \\ \\ \\ \\ \\ \\
        
        \multirow{3}{*}{Transmission Capacity} & \multirow{5}{*}{\parbox{8cm}{The extent of grid reinforcement in Europe beyond 2030 remains uncertain.
        However, future changes are likely to be minor compared to the existing network layout.}} & \multirow{5}{*}{\parbox{5 cm}{We consider the impact of a 12.5$\%$ and 25$\%$ expansion in additional volume, measured in GWkm, on EGS capacity.}} \\ \\ \\ \\ \\

        \multirow{4}{*}{Waste Heat} & \multirow{5}{*}{\parbox{8cm}{Fuel synthesis processes, such as Fischer-Tropsch, are anticipated to provide a low-cost source of district heating. However, the extent of this supply remains uncertain.
        For heat (co-)generating \gls{egs}, we assess the impact of excluding Fischer-Tropsch as a low-cost heat source.}} & \multirow{5}{*}{\parbox{5cm}{The utilisation of waste heat from fuel synthesis for low-temperature heat demands is deactivated.}} \\ \\ \\ \\ \\

        \multirow{4}{*}{Renewable Cost} & \multirow{5}{*}{\parbox{8cm}{The costs of wind, solar, and hydro generators have been updated.}} & \multirow{5}{*}{\parbox{5cm}{A 20\% increase or decrease in the base case capital expenditure.}} \\ \\ \\ \\ \\

        \multirow{4}{*}{District Heating} & \multirow{5}{*}{\parbox{8cm}{Although the expansion of district heating in Europe is included in many energy transition outlooks, its future rollout remains uncertain.
        PyPSA-Eur models varying levels of rollout using a parameter, where 0 represents the 2020 rollout level, and 1 corresponds to all regions meeting 60\% of their urban residential heat demand through district heating.
        In the assumed base case with a rollout progress of 0.3, each region is scaled individually.}} & \multirow{5}{*}{\parbox{5cm}{A zero-progression scenario assumes district heating shares remain at 2020 levels.
        Additionally, two scenarios are considered where district heating progresses to 60\% and 100\% of urban residential heat demand, respectively.}} \\ \\ \\ \\ \\
 
    \end{tabular}
    \label{tab:sensitivities}
\end{table*}


\newpage

\makeatletter
\renewcommand \thesection{S\@arabic\c@section}
\renewcommand\thetable{S\@arabic\c@table}
\renewcommand \thefigure{S\@arabic\c@figure}
\makeatother
\renewcommand{\citenumfont}[1]{S#1}
\setcounter{equation}{0}
\setcounter{figure}{0}
\setcounter{table}{0}
\setcounter{section}{0}




\end{document}